\documentclass[a4paper,10pt]{article}
\usepackage{graphicx}
\usepackage{amsmath}
\usepackage{amssymb}
\usepackage{hyperref}
\usepackage{enumitem}
\usepackage{mathtools}

\usepackage{amsthm}

\usepackage{siunitx} 
\usepackage[numbers]{natbib}

\usepackage{multicol} 
\usepackage{float}

\usepackage{titlesec}

\titleformat{\section}{\normalfont\Large\bfseries\MakeUppercase}{\thesection}{1em}{}
\titleformat{\subsection}{\normalfont\large\bfseries\MakeUppercase}{\thesubsection}{1em}{}
\titleformat{\subsubsection}{\normalfont\normalsize\bfseries\MakeUppercase}{\thesubsubsection}{1em}{}

\usepackage[top=1in, bottom=1in, left=0.75in, right=0.75in]{geometry}

\usepackage{helvet}

\usepackage{fancyhdr}
\makeatletter
\patchcmd{\@maketitle}{\thispagestyle{plain}}{\thispagestyle{fancy}}{}{}
\makeatother

\fancypagestyle{plain}{ 
  \fancyhf{}
  \fancyhead[L]{\url{https://doi.org/10.48550/arXiv.2408.09218}}
  \fancyfoot[C]{\thepage}
}

\makeatletter
\renewenvironment{abstract}{
    \small
    \noindent\textbf{\textit{\abstractname}}.\hspace{1em}} 
    {}
\makeatother

\usepackage[parfill]{parskip} 
\usepackage{setspace}

\title{\textbf{\underline{FQGA-single}: Towards Fewer Training Epochs and Fewer Model Parameters for Image-to-Image Translation Tasks}\\[0.5em]
\rule{\textwidth}{0.7pt}}

\date{}
\author{
    {Russell Cho Yang} \\
    School of Physical and Mathematical Sciences,
    Nanyang Technological University, Singapore
}

\begin{document}

\maketitle

\begin{abstract}
The utilization of 3D cone-beam computed tomography (CBCT) in cancer treatment, particularly within image-guided adaptive radiation therapy (IGART) workflows for photon and proton therapy, is integral \cite{Letourneau}, \cite{Guckenberger}. However, the effectiveness of CBCT is hindered by image artifacts \cite{Kim}, \cite{Stock}, necessitating the introduction of synthetic CT (sCT) to elevate CBCT quality to the level of conventional CT scans. This enhancement is crucial for accurate dose calculations and online adaptive workflows in radiation therapy, ultimately enhancing IGART quality for patients. This paper proposes a novel model inspired by \textbf{CycleGAN} \cite{Zhu}: \textbf{FQGA-single} to produce high quality sCT images even more efficiently. Evaluations were done on SynthRAD Grand Challenge dataset with CycleGAN model used for benchmarking and for comparing the quality of CBCT-to-sCT generated images from a quantitative and qualitative perspective. Finally, this paper also explores ideas from the paper \textbf{"One Epoch Is All You Need"} \cite{Komatsuzaki} to compare models trained on a single-epoch versus multiple-epochs or multi-epochs. Astonishing results from FQGA-single were obtained during this exploratory experiment which shows performance of FQGA-single when trained on single-epoch surpassing itself when trained on multi-epoch. More surprising is its performance also surpassing CycleGAN's multi-epoch and single-epoch models, and even a modified version of CycleGAN, \textbf{CycleGAN-m} in Section \textbf{9}.

We provide TensorFlow implementation of this paper \href{https://github.com/choyang002/FQGA-single}{\textbf{\textcolor{red}{here}}}

This research was carried out together with researchers and medical clinicians at the National Cancer Center Singapore.
\end{abstract}


\section{Introduction}
An AI Model which is fast to train and lightweight is an effort towards an ambitious goal — to democratize technology. AI often requires huge computational and time resources, making it inaccessible for certain groups
of people who may lack these resources. Moreover, certain more advanced technologies have yet to be more broadly deployed in areas which requires such technology but are not able to implement them due to their limited technological hardware capabilities. Therefore, this paper aims to find a fast and good model which performs well in Image-to-Image translation tasks.

Medical CBCT-CT image translation is a unique problem as the performance of its algorithm is dependent on both its quantitative and qualitative performance. Quantitative performance can be measured using metrics like PSNR, SSIM, MAE and MSE while qualitative performance may be more subjective in nature although still equally as important. Ideally, we would like to develop a CBCT-CT translation algorithm which is able to perform well both in terms of quantitative and qualitative aspects. However, as we see in this paper, we notice that good quantitative performance does not necessarily mean good qualitative performance and good quantitative performance does not necessarily mean good qualitative performance.

Therefore, to produce an algorithm that not only had better qualitative
performance but better quantitative performance as well, \textbf{FQGA (Fast Paired Image-to-Image Translation Quarter-Generator Adversary) Model} is proposed to reduce visual artifacts (e.g. smudging of pixels within sCT generated images and streaks of pixel noise on sCT generated images). FQGA is a lightweight model which has $\frac{1}{4}$ the number of parameters for its generator model as compared to current State-Of-The-Art (SOTA) CycleGAN model \cite{Zhu}.

With inspiration from training language models with a single epoch \cite{Komatsuzaki}, this paper also proposes a data pipeline for single-epoch training called Single-Epoch Modification (SEM) method. CycleGAN model \cite{Zhu} was trained on SynthRAD Grand Challenge Dataset using the SEM method, we called (CycleGAN-single) in this paper. Versus (CycleGAN-multi) which refers to training CycleGAN model for 200 epochs like most papers. Although qualitative performance of CycleGAN-single was better than that of CycleGAN-multi, its quantitative performance was not as good.

\section{Objective and Scope}

The objectives of this paper are as follows:

\begin{enumerate}

    \item \textbf{SynthRad Dataset Overview}: Overview of the dataset
and conversion from 3D CBCT-CT paired volume data
of every patient into 2D CBCT-CT paired slice data.

    \item \textbf{Training Data Pipeline}: comparisons with standard
training workflow of CycleGAN for image-to-image
translation tasks. Multi-Stage Padding: Algorithm for
ideal 2D slice Reflection and Constant padding to
preserve border and output features.

    \item \textbf{CycleGAN Performance Comparison}: CycleGAN-single versus CycleGAN-multi from a Quantitative and Qualitative Performance Comparison.

    \item \textbf{FQGA Model Proposal}: FQGA Architecture.

    \item \textbf{FQGA and CycleGAN Performance Comparison.}

    \item \textbf{FQGA Ablation Studies.}

    \item \textbf{FQGA Performance with SEM Method.}
    
\end{enumerate}

\section{SynthRad Dataset Overview}

\subsection{Data Cleaning}
The SynthRAD Grand Challenge dataset consists of
180 CBCT-CT paired volume data. For more info, one
can refer to the following link
(\url{https://doi.org/10.5281/zenodo.7260704}) on the unique
image acquisition parameters for each patient data (e.g.
machine model, slice thickness) for each 3D CBCT-CT
paired volume patient data. After KDE-Based
Classification, 30 CBCT-CT paired volume data is
discarded and only 150 CBCT-CT paired volume data
remain. Each CBCT or CT volume has around 50 to
105 slices.

\subsection{Training, Validation and Test Data}
After data cleaning, there are 150 volume data pairs
from 3 data centers, each data center contributing
around 50 volume data pairs. From these 150 paired
volume data, we performed 2 experiments. Experiment
1) is the experiment which will be detailed in this paper
and involves the usage of 35 CBCT-CT paired volume
data for training, 15 paired volume data for validation
and 10 paired volume data for testing. In Experiment 2),
we further show
that CycleGAN-single outperforming CycleGAN-multi
is not an anomalous result which is only the case for a
specially selected 10 paired volume testing data. But,
even with the use of the remaining SynthRAD dataset
consisting of 100 paired volume data for testing, this
result still holds. (100 paired volume data derived when
original 150 paired volume data minus 35 paired
volume data for training and minus 15 paired volume
data for validation)

\subsection{Uniformity of Data}
In SynthRAD Grand Challenge Dataset and in many
other real-world datasets, there is noise in the dataset
and more importantly, dimensions of the 3D CBCT-CT
volume pair data vary due to the uniqueness of every
patient data. In our implementation, we converted the
3D CBCT-CT paired volume data (Height, Width and
Depth) originally in SynthRAD dataset into 2D CBCTCT paired data (Height and Width). To attain uniformity in the dimensions of 2D input data being fed into the model, the largest Height and Width
dimensions of input of every training batch is stored
and all other 2D images in the training data are padded
using Reflection Padding to be equals to those
dimensions. Moreover, to ensure that the resolution of
the output image of the generator model is equal to its
input image, there was Constant Padding done to ensure that images are padded to a resolution that is a factor of 4 (This is to accommodate the down-sample and upsample procedures of the generator model). Afterwards,
Hounsfield Units (HU) values in both CBCT and CT slices are normalized to be of values between -1 and 1 for training, validation, and testing.

\section{Training Data Pipeline}
The novelty in this workflow lies in 3 areas. First, the
Reflection padding done in Section 1 is not usually done in the standard workflow of Generative Adversarial Network (GANs) workflow for image-toimage translation problems. Next, in standard GANs workflow namely CycleGAN, Reflection Padding is usually done between the layers of CycleGAN
Generator and Discriminator models. In our implementation, instead of Reflection Padding, Constant Padding was used. Moreover, we introduced a
dynamic Constant Padding algorithm which was able to identify the optimal padding for each input and output layer of our FQGA models and therefore, the term “Multi-Stage Padding”.

\section{CycleGAN Performance Comparison}

\subsection{Quantitative Test Performance}

\textbf{*CycleGAN-multi is the performance after 200 epochs.}

\textbf{*CycleGAN-multi-1 is the performance after 1 epoch.}

\begin{figure}[H]
    \centering
    \includegraphics[width=0.7\linewidth]{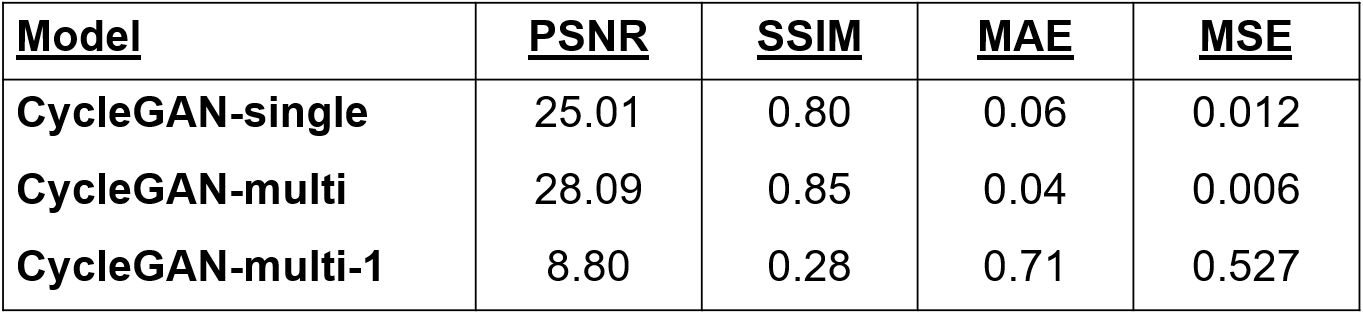} 
    \caption{CycleGAN-single vs CycleGAN-multi}
    \label{fig:1}
\end{figure}

From Section 5.1, we can see that there is significant
improvement in quantitative performance when
CycleGAN is trained using the SEM Method. (More information about SEM Method found in Appendix, Section 14.)
This is different from CycleGAN-multi-1 which is trained on a single epoch
without using SEM Method.
Moreover, in Section 5.2, we also can see that the
generated images from CycleGAN-single are similar to
those of CycleGAN-multi. While CycleGAN-multi-1
generated images which did not even have the outline of
the object. Therefore, although quantitatively,
CycleGAN-single showed an improvement, generated
qualitative images are still poor and CycleGAN-single
quantitaive perfomance is still less than that of
CycleGAN-multi.

\subsection{Qualitative Test Performance}

\begin{figure} [H]
    \centering
    \includegraphics[width=0.85\linewidth]{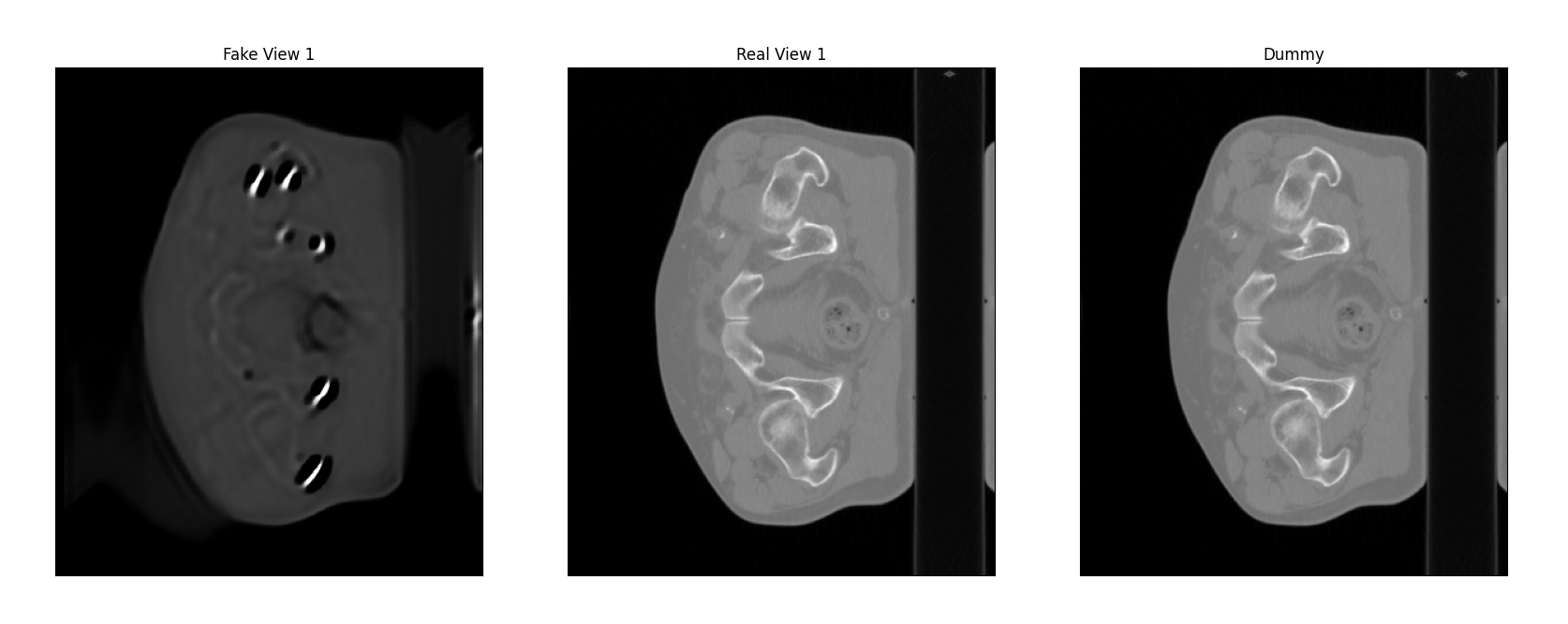}
    \caption{CycleGAN-single (Generated Image 1)}
    \label{fig:2}
\end{figure}

\begin{figure} [H]
    \centering
    \includegraphics[width=0.85\linewidth]{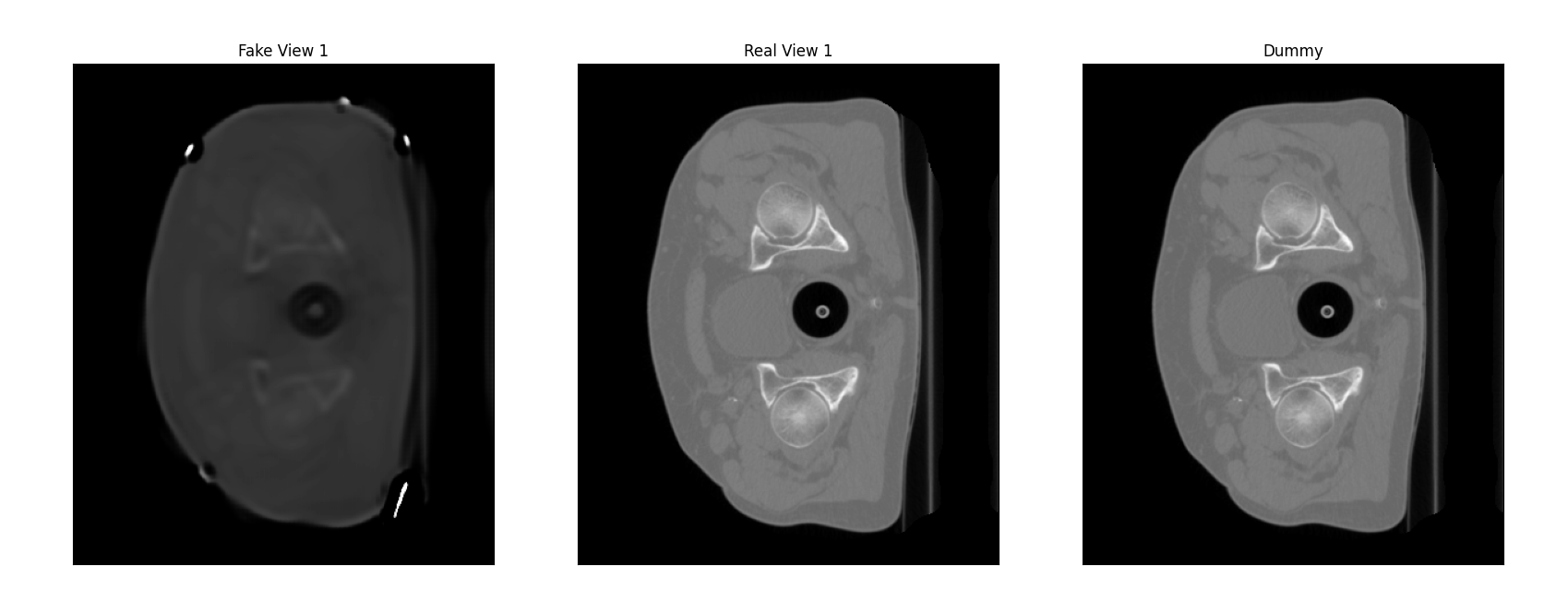}
    \caption{CycleGAN-single (Generated Image 2)}
    \label{fig:3}
\end{figure}

\begin{figure} [H]
    \centering
    \includegraphics[width=0.85\linewidth]{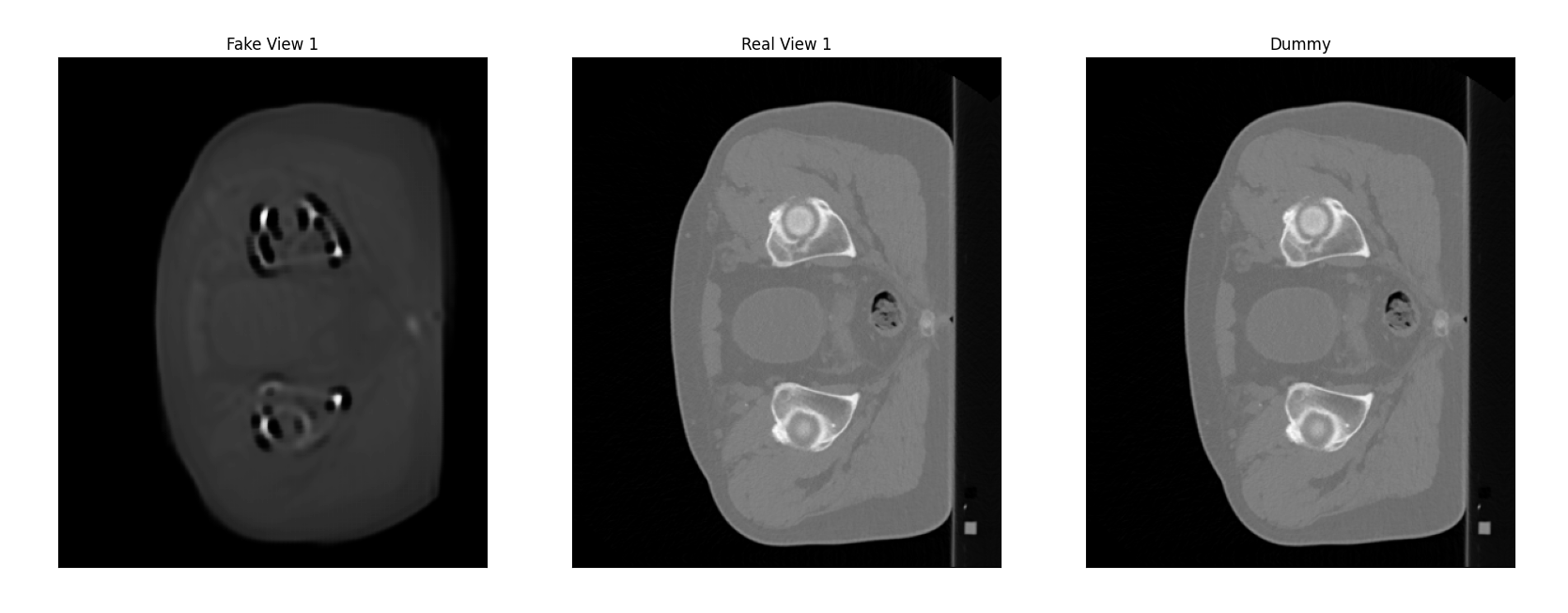}
    \caption{CycleGAN-single (Generated Image 3)}
    \label{fig:4}
\end{figure}

\begin{figure} [H]
    \centering
    \includegraphics[width=0.85\linewidth]{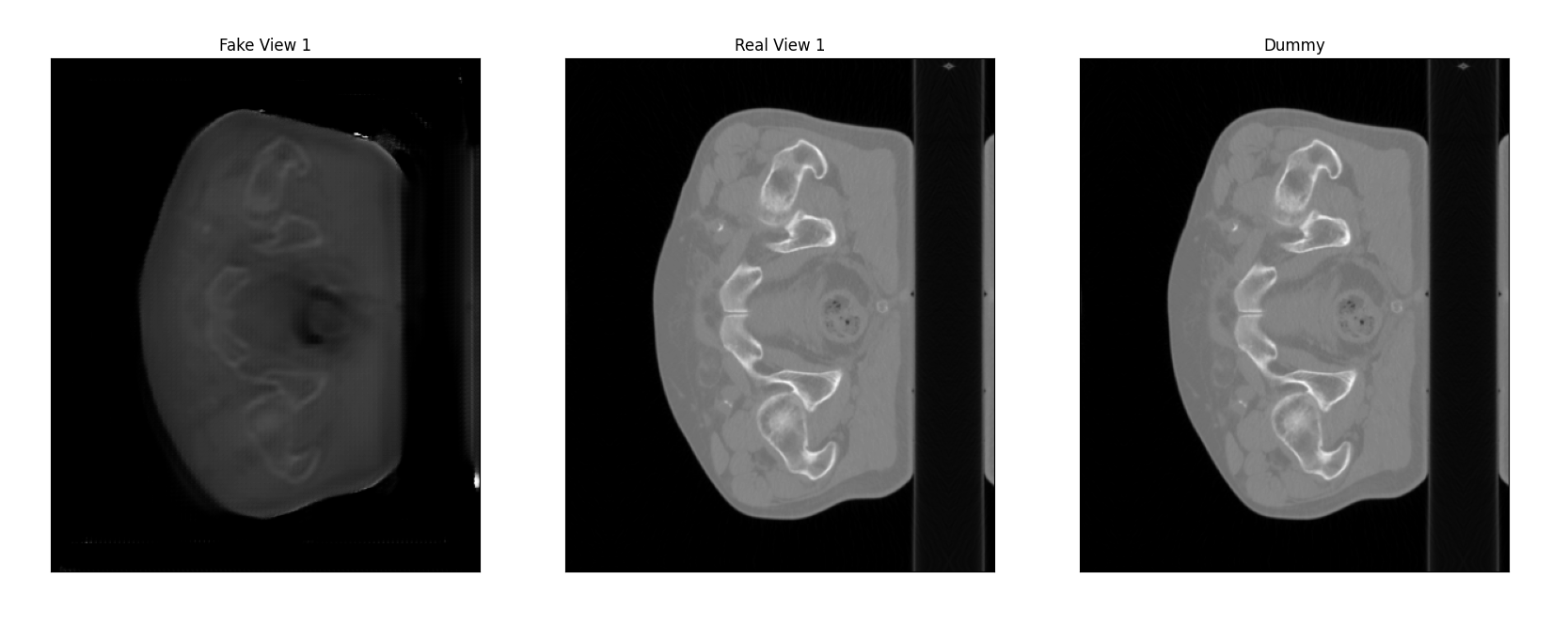}
    \caption{CycleGAN-multi (Generated Image 1)}
    \label{fig:5}
\end{figure}

\begin{figure} [H]
    \centering
    \includegraphics[width=0.85\linewidth]{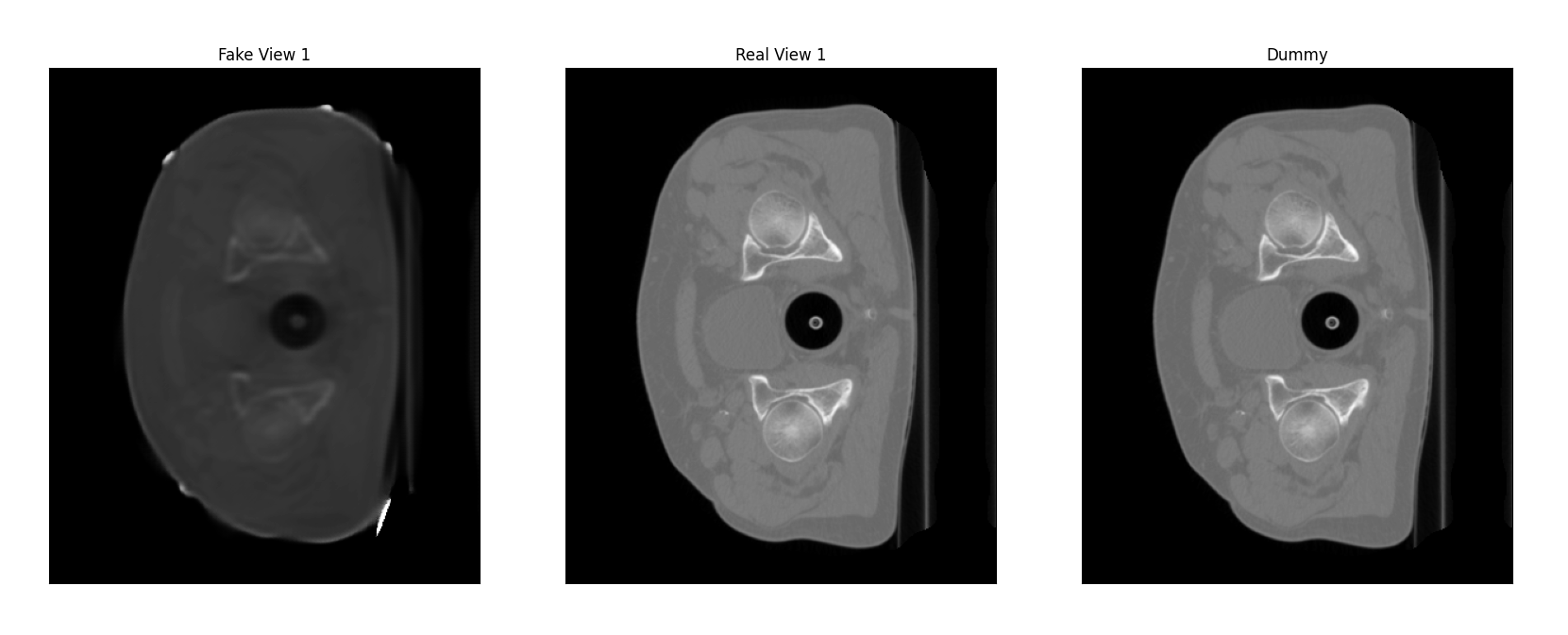}
    \caption{CycleGAN-multi (Generated Image 2)}
    \label{fig:6}
\end{figure}

\begin{figure} [H]
    \centering
    \includegraphics[width=0.85\linewidth]{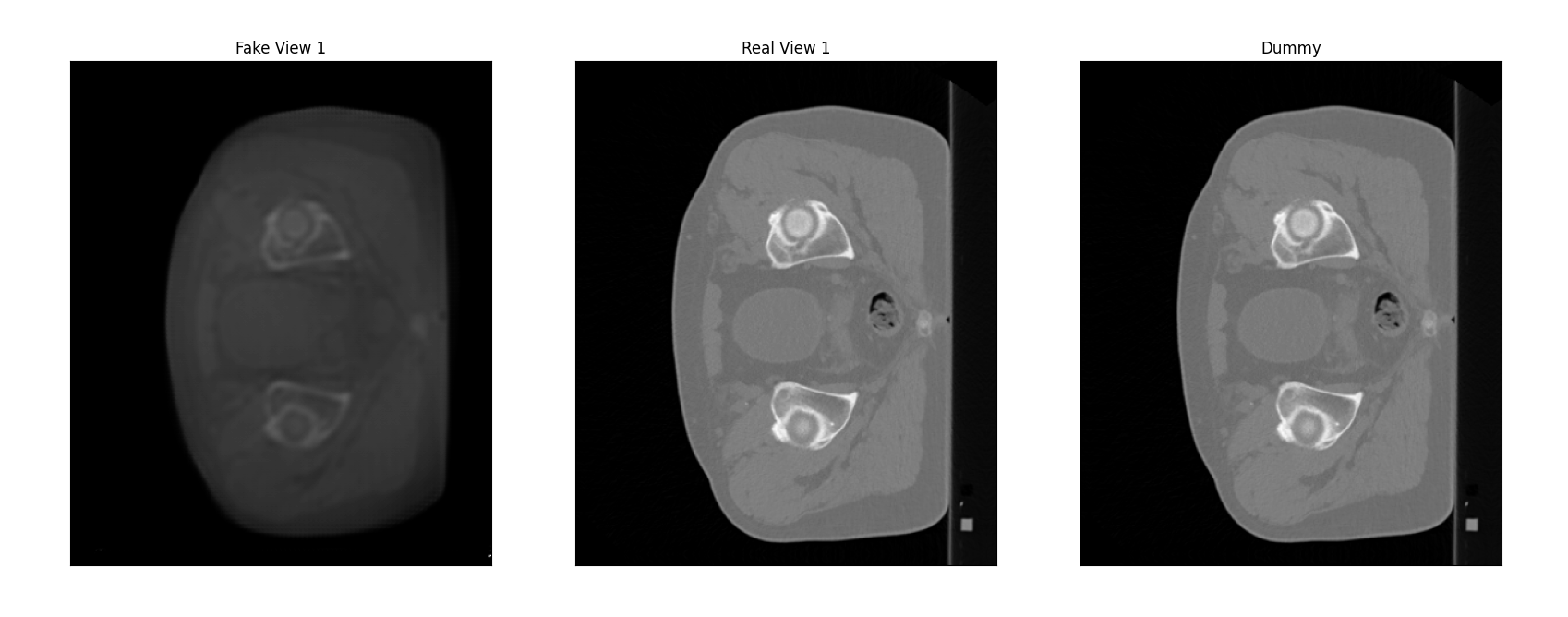}
    \caption{CycleGAN-multi (Generated Image 3)}
    \label{fig:7}
\end{figure}

\begin{figure} [H]
    \centering
    \includegraphics[width=0.85\linewidth]{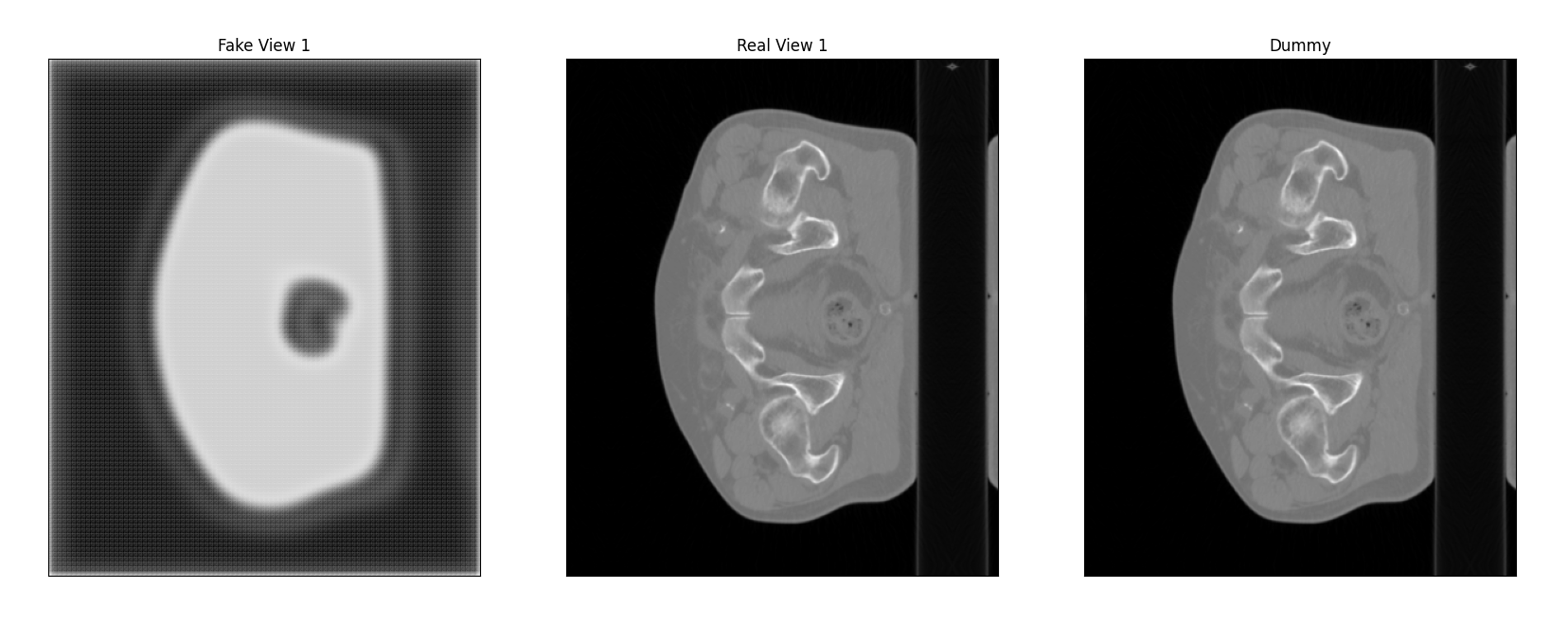}
    \caption{CycleGAN-multi-1 (Generated Image 1)}
    \label{fig:8}
\end{figure}

\begin{figure} [H]
    \centering
    \includegraphics[width=0.85\linewidth]{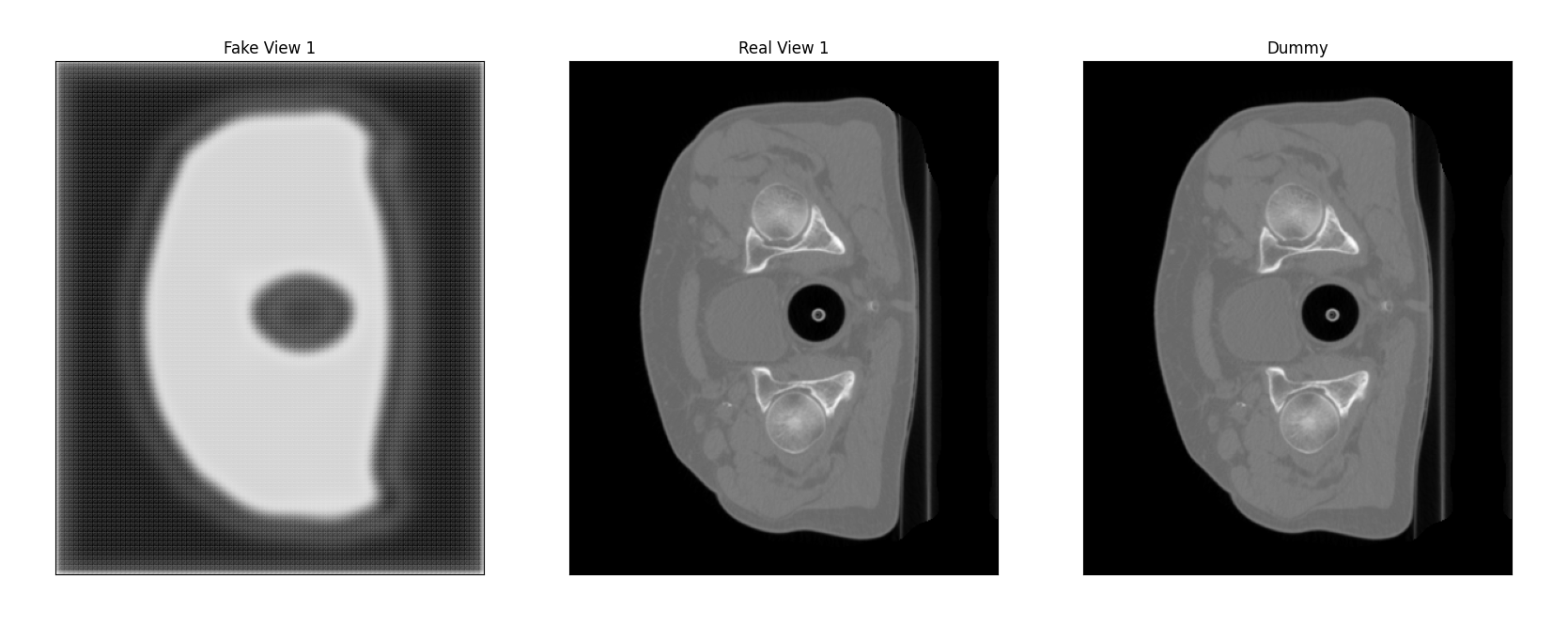}
    \caption{CycleGAN-multi-1 (Generated Image 2)}
    \label{fig:9}
\end{figure}

\begin{figure} [H]
    \centering
    \includegraphics[width=0.85\linewidth]{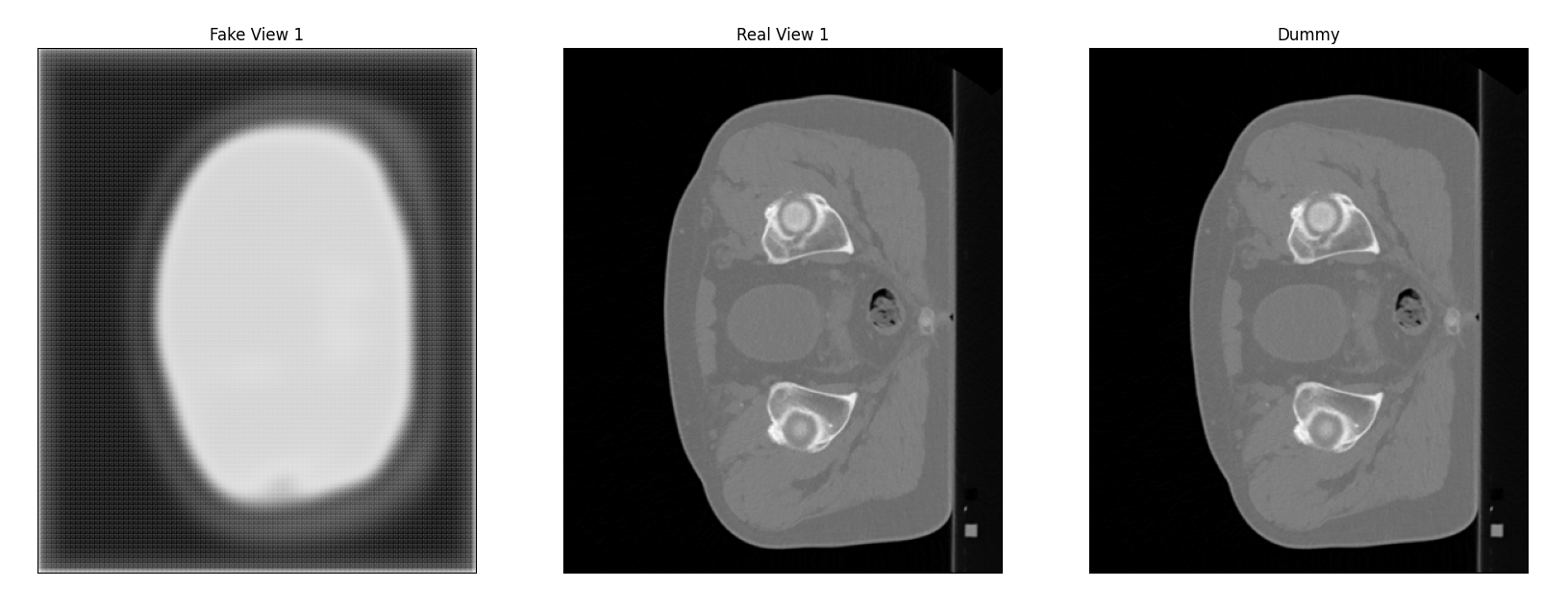}
    \caption{CycleGAN-multi-1 (Generated Image 3)}
    \label{fig:10}
\end{figure}

In Section 6, we propose a FQGA model
which is able to outperform CycleGAN-multi in terms
of its quantitative performance, while preserving if not
improving the qualitative performance of CycleGAN-single. Finally, in Section 9, we trained FQGA
model using the SEM method for fast training and
evaluated its surprisingly good performance which surpassed CycleGAN.

\section{FQGA Model Proposal}

\subsection{FQGA Discriminator Model Architecture Overview}


\begin{figure}[H]
    \centering
    \includegraphics[height=0.9\textheight, keepaspectratio]{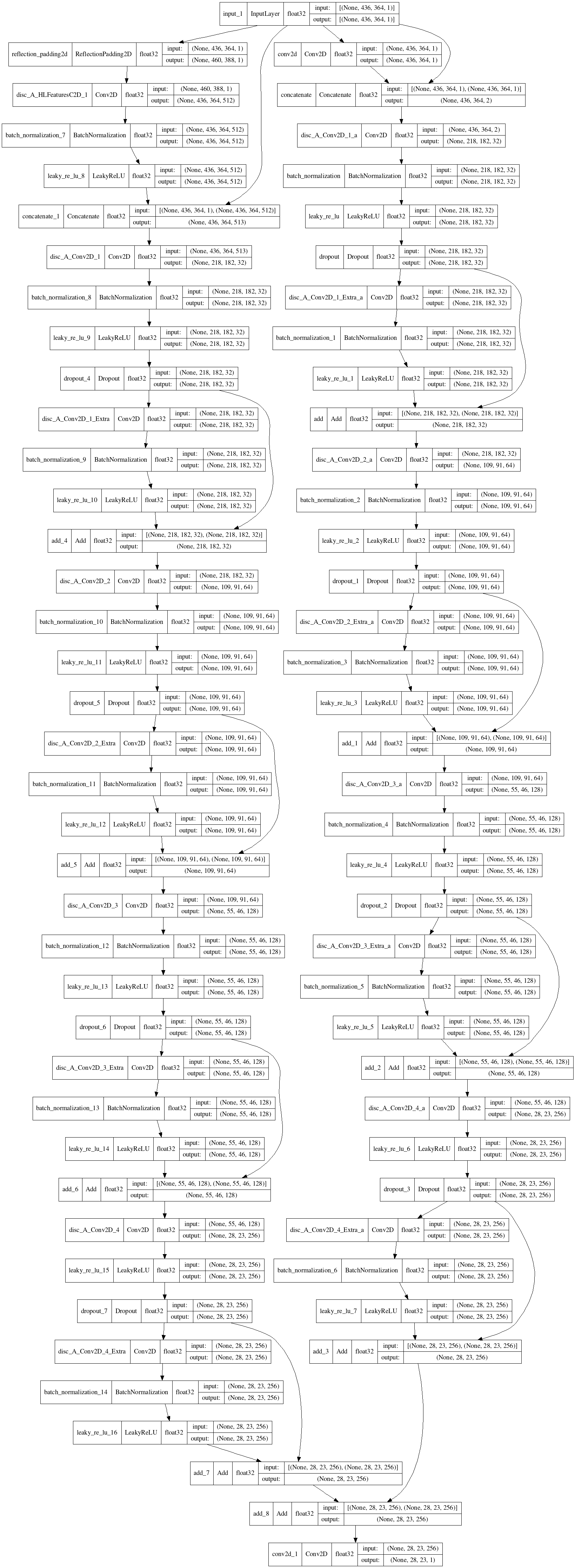}
    \caption{FQGA Discriminator Model Architecture}
    \label{fig:50}
\end{figure}

\subsection{FQGA Discriminator Model Description}
There are 2 parallel inputs. For the first parallel input,
as the initial (original) input image is fed into the
discriminator model, a Gaussian blur kernel of [[1/16,
1/8, 1/16], [1/8, 1/4, 1/8], [1/16, 1/8, 1/16]] is used to
blur the image. The original image is then concatenated
with the original image along the last dimension. These
concatenated output is then passed to a 2D
Convolutional Layer A1 of size (2x2) with strides (2x2),
number of filters=32 and Dropout. The output of this
Convolutional Layer is then fed into a 2D
Convolutional Layer B1 of size (3x3) with strides (1x1)
and number of filters=32. The output of Convolutional
Layer A1 and Convolutional Layer B1 are then Added
together before feeding to the next 2D Convolutional
Layer A2 of size (3x3) with strides (2x2), number of
filters=64 and Dropout. The output of this
Convolutional Layer is then fed into a 2D
Convolutional Layer B2 of size (5x5) with strides (1x1)
and number of filters=64. There are 4 of such
Convolutional Layer pairs Ax and Bx where x=[1,4]
with increasing filter sizes and increasing number of
filters. In this block of layers, Batch Normalization and
LeakyReLU activation are used. For the second parallel
input, the initial input image is fed into the
discriminator model, a (25x25) Constant Padding is
applied to the image before being fed into a 2D HighLevel Convolutional Layer of size (25x25) with strides
(1x1) and number of filters=512. The output of this
High-Level Convolutional Layer is concatenated with
the original image along the last dimension. Afterwards,
similar to the first parallel input, the output are then fed
into 4 Convolutional Layer pairs Ax and Bx where
x=[1,4]. Finally, the output of these 2 parallel inputs are
Added together and passed to a (10x10) Patch Output.
Refer to the Appendix Section for more details.

\clearpage
\subsection{FQGA Generator Model Architecture Overview}


\begin{figure}[H]
    \centering
    \includegraphics[height=0.9\textheight, keepaspectratio]{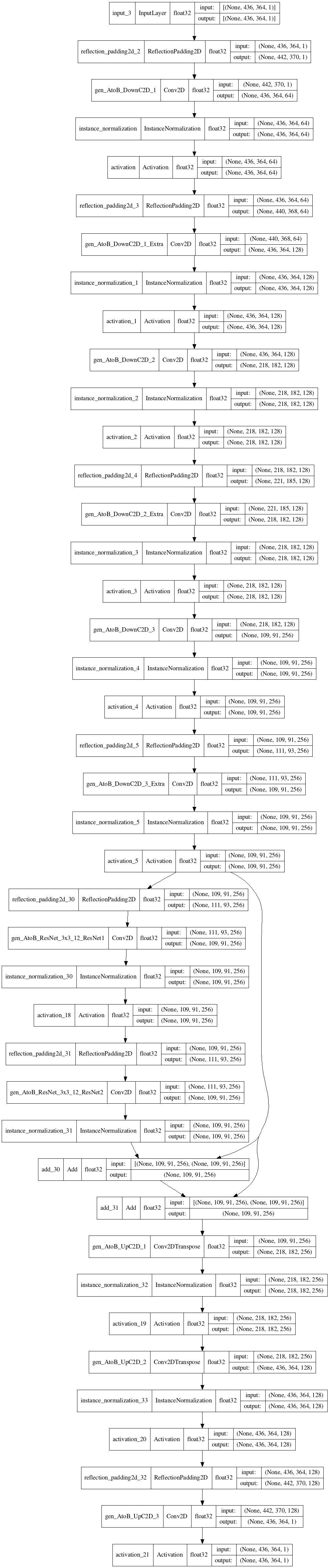}
    \caption{FQGA Generator Model Architecture}
    \label{fig:51}
\end{figure}

\subsection{FQGA Generator Model Description}
The number of parameters for FQGA discriminator
model is greater than that of CycleGAN’s but I
hypothesized that the number of parameters for the
discriminator in GANs should actually be more than
that of the generator model so that the discriminator
model can act as a stronger form of ground-truth for the
generator model to optimize its model accordingly in
the correct direction with less noise. Moreover, when
generating images, the generator model is usually the
only model required while the discriminator model has
to be present during the training process of the GANs.
In FQGA, its Generator Model, has only ¼ the number
of parameters compared to CycleGAN. In FQGA
generator model, there are two CNN Layers of sizes
(7x7) and (5x5) which have the number of filters 64 and
128 where both of these layers have strides of (1x1).
Next is the downsampling block of Convolutional
Layers of (4x4) and (3x3) with strides (2x2) and
Convolutional Layers of (4x4) and (3x3) with strides
(1x1) in between the downsampling CNN layers of the
downsampling block. In the bottleneck layer, there is
only 1 FQGA-layer or FQGA-skip-layer which is inspired by residual-blocks \cite{He}. However, instead of in the CycleGAN implementation where there are originally 9 residual
blocks \cite{He}, for FQGA-single, there is effectively only 1 residual block for the bottleneck layer. After the bottleneck layer, FQGA has upsampling block of two
(4x4) filter sizes with number of filters reducing from
256 to 128 and finally, a hyperbolic tangent activation
function. FQGA model is trained with Adam
optimization with a batch size of 1. All weights are
initialized from random normal initializer with a mean
of 0 and a standard deviation of 0.02. Instance
Normalization is used after convolution, and
LeakyReLU with slope 0.2 is used after instance
normalization for fast convergence \cite{Wang}.

\clearpage
\section{FQGA and CycleGAN Performance Comparison}

\subsection{Quantitative Test Performance}
\textbf{*FQGA-single-20 is the performance after 20 epochs.}

\begin{figure} [H]
    \centering
    \includegraphics[width=0.7\linewidth]{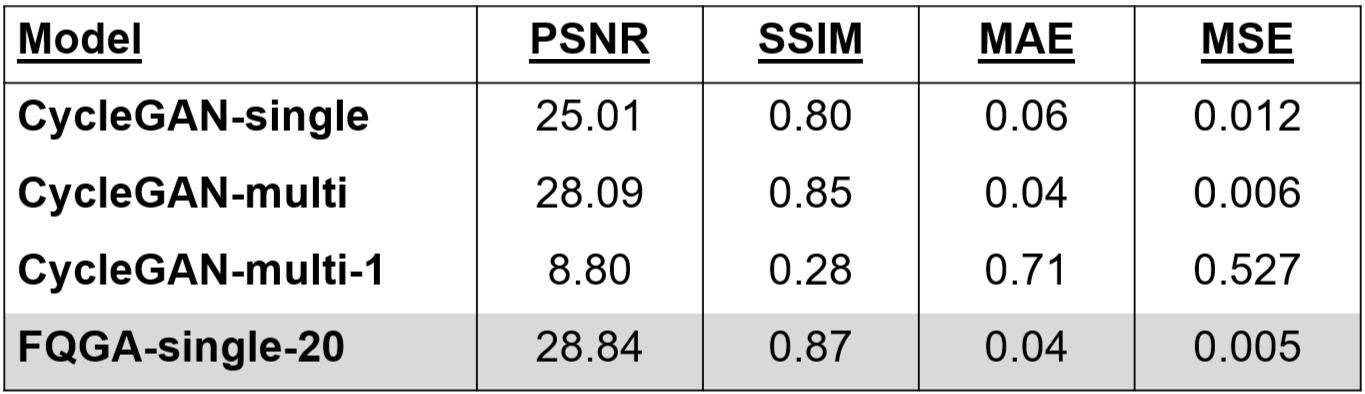}
    \caption{FQGA vs CycleGAN}
    \label{fig:11}
\end{figure}

\subsection{Qualitative Test Performance}

\begin{figure} [H]
    \centering
    \includegraphics[width=0.85\linewidth]{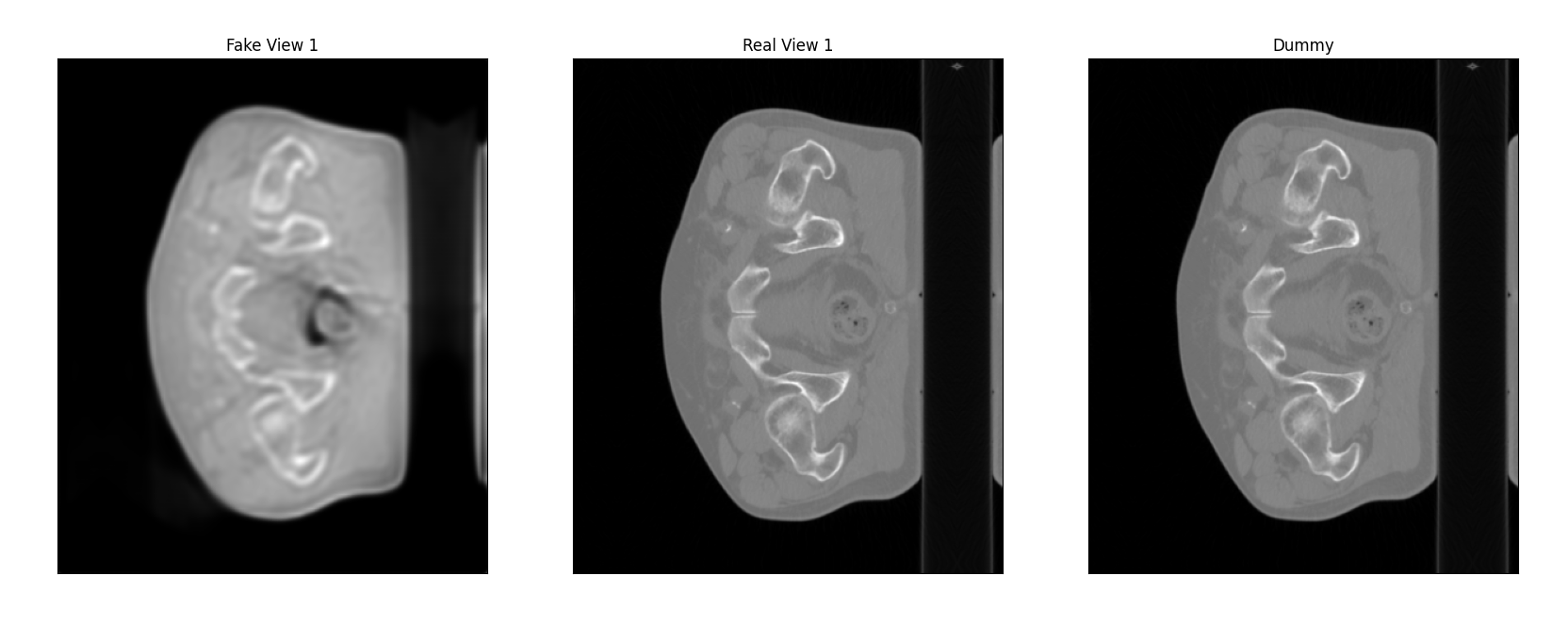}
    \caption{FQGA-single-20 (Generated Image 1)}
    \label{fig:12}
\end{figure}

\begin{figure} [H]
    \centering
    \includegraphics[width=0.85\linewidth]{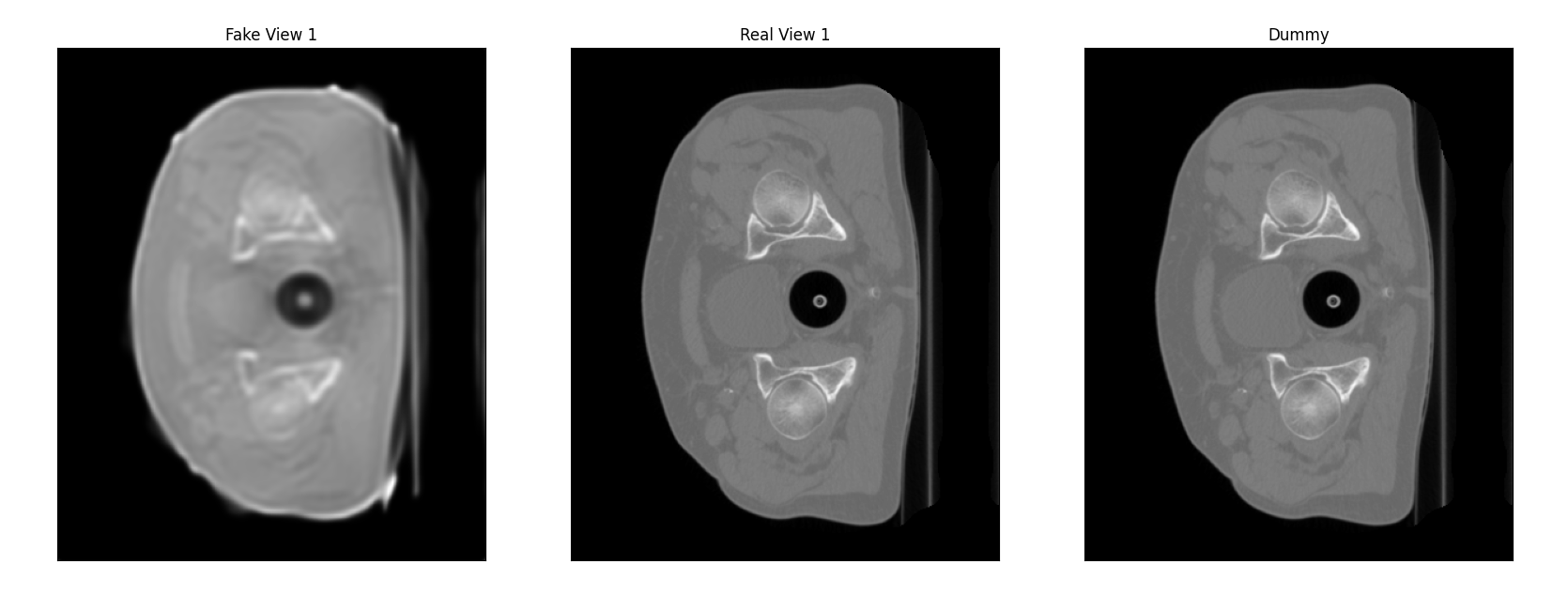}
    \caption{FQGA-single-20 (Generated Image 2)}
    \label{fig:13}
\end{figure}

\begin{figure} [H]
    \centering
    \includegraphics[width=0.85\linewidth]{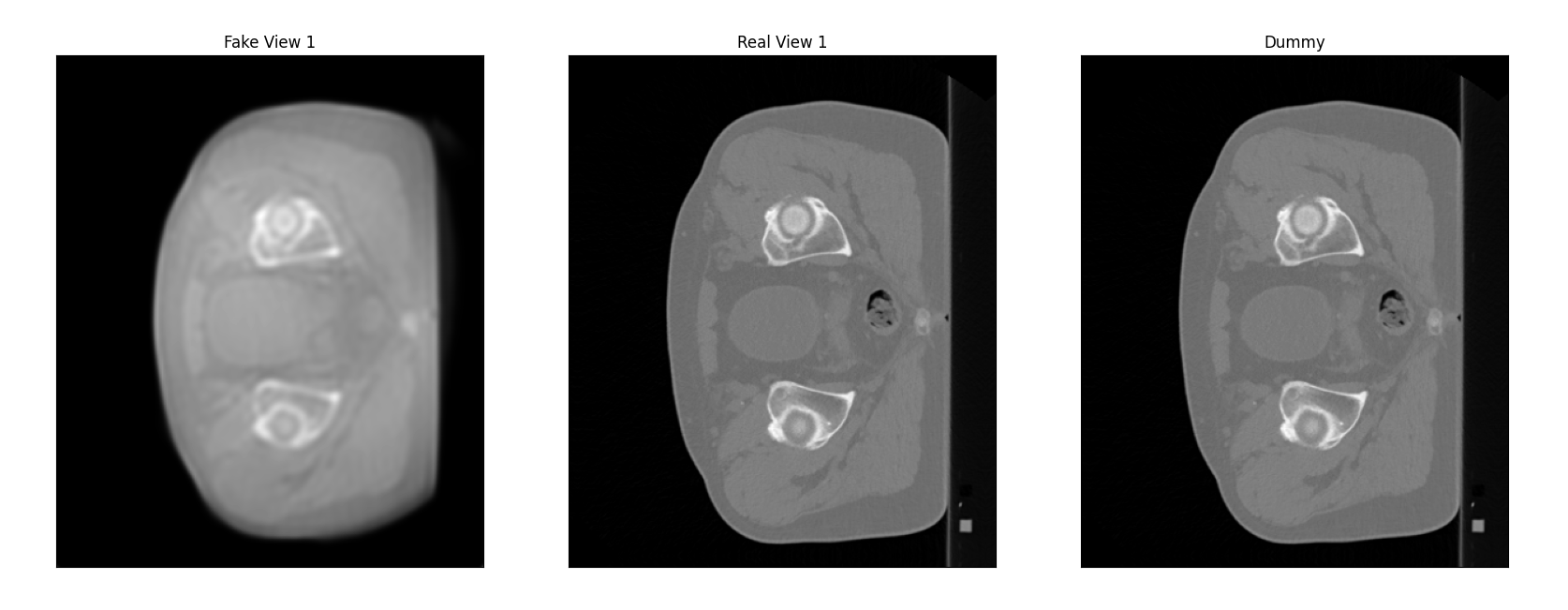}
    \caption{FQGA-single-20 (Generated Image 3)}
    \label{fig:14}
\end{figure}

\section{FQGA Ablation Studies}

\subsection{Quantitative Test Performance}
\textbf{*CycleGAN-1Res is when vanilla CycleGAN has 1 ResNet [7] block
instead of 9. This model is trained on 200 epochs.}

\textbf{*CycleGAN-Disc-20 is when vanilla CycleGAN’s Discriminator
Model is replaced with FQGA’s Discriminator Model. This model is
trained on 20 epochs.}

\textbf{*CycleGAN-Gen-20 is when vanilla CycleGAN’s Generator Model is
replaced with FQGA’s Generator Model. This model is trained on 20
epochs.}

\begin{figure} [H]
    \centering
    \includegraphics[width=0.7\linewidth]{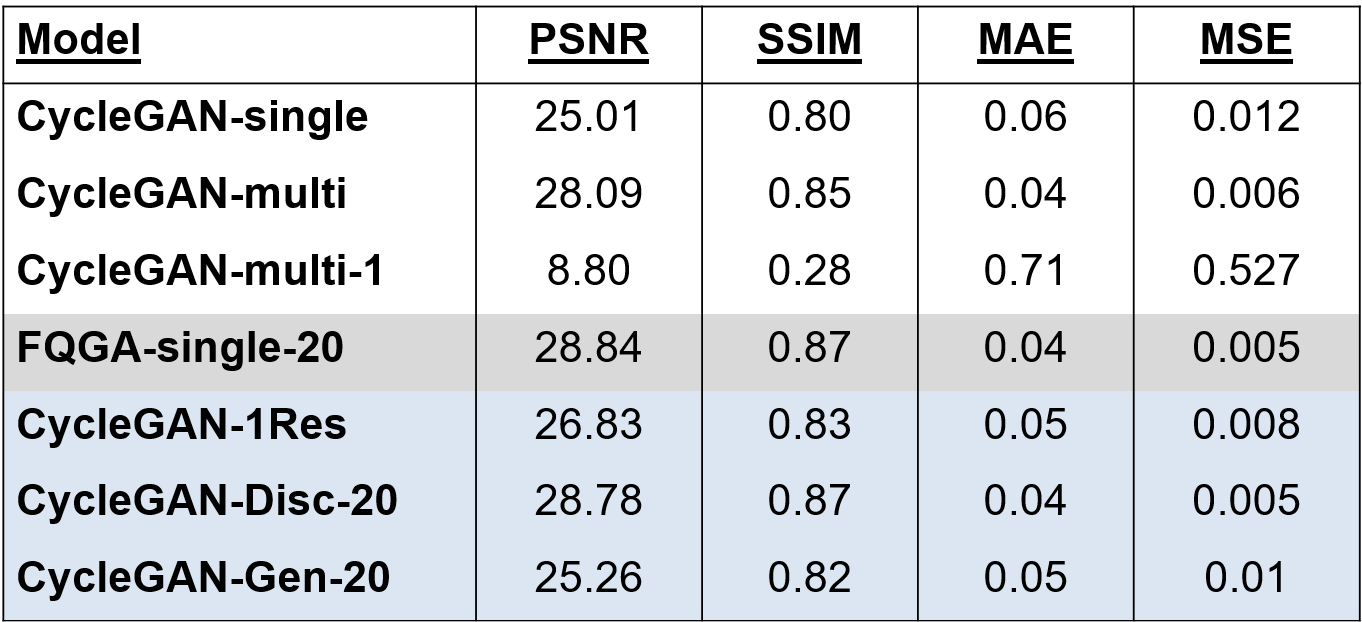}
    \caption{FQGA Ablation Studies}
    \label{fig:15}
\end{figure}

\subsection{Qualitative Test Performance}

\begin{figure} [H]
    \centering
    \includegraphics[width=0.85\linewidth]{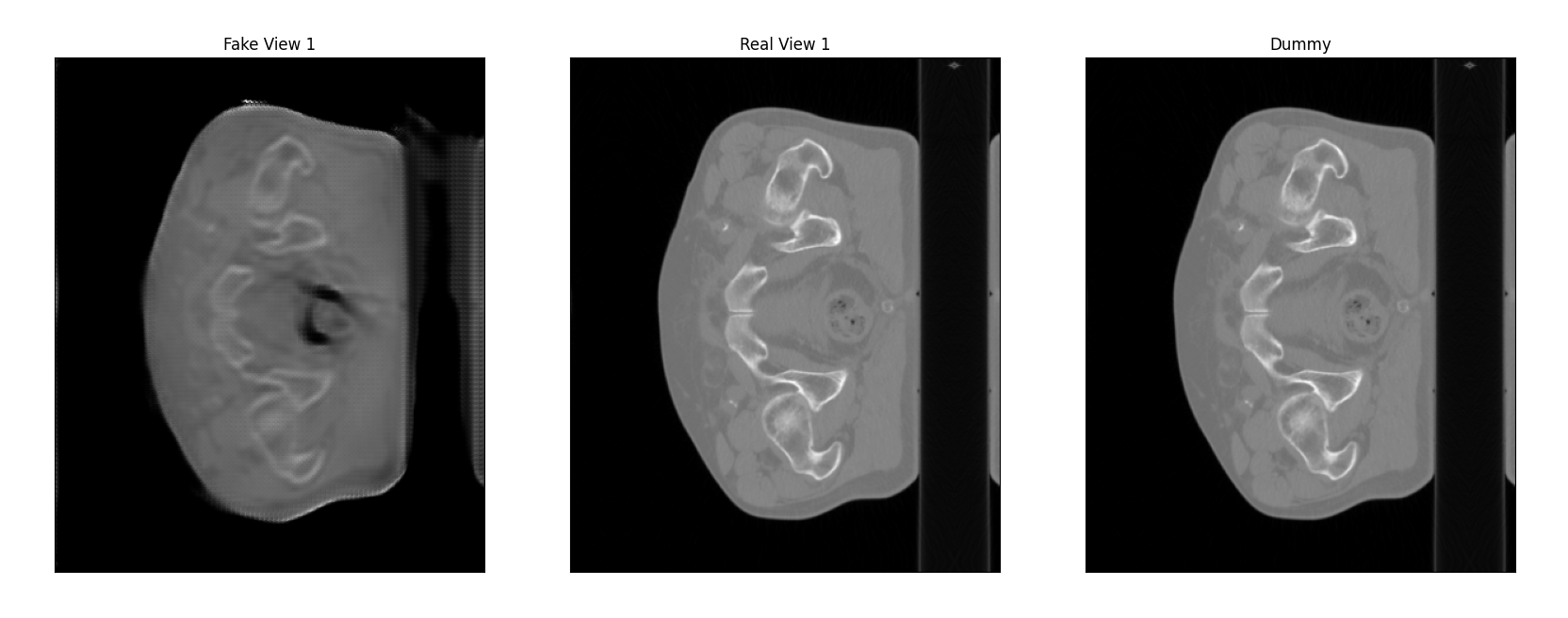}
    \caption{CycleGAN-1Res (Generated Image 1)}
    \label{fig:16}
\end{figure}

\begin{figure} [H]
    \centering
    \includegraphics[width=0.85\linewidth]{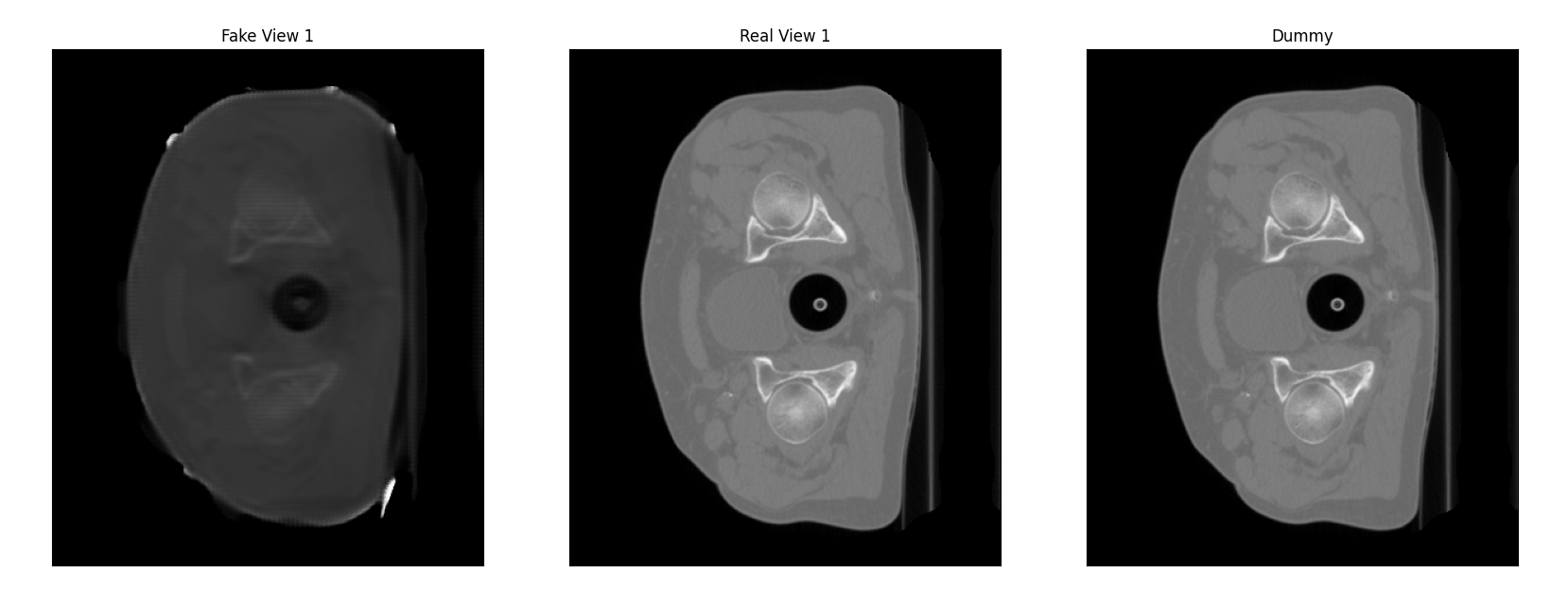}
    \caption{CycleGAN-1Res (Generated Image 2)}
    \label{fig:17}
\end{figure}

\begin{figure} [H]
    \centering
    \includegraphics[width=0.85\linewidth]{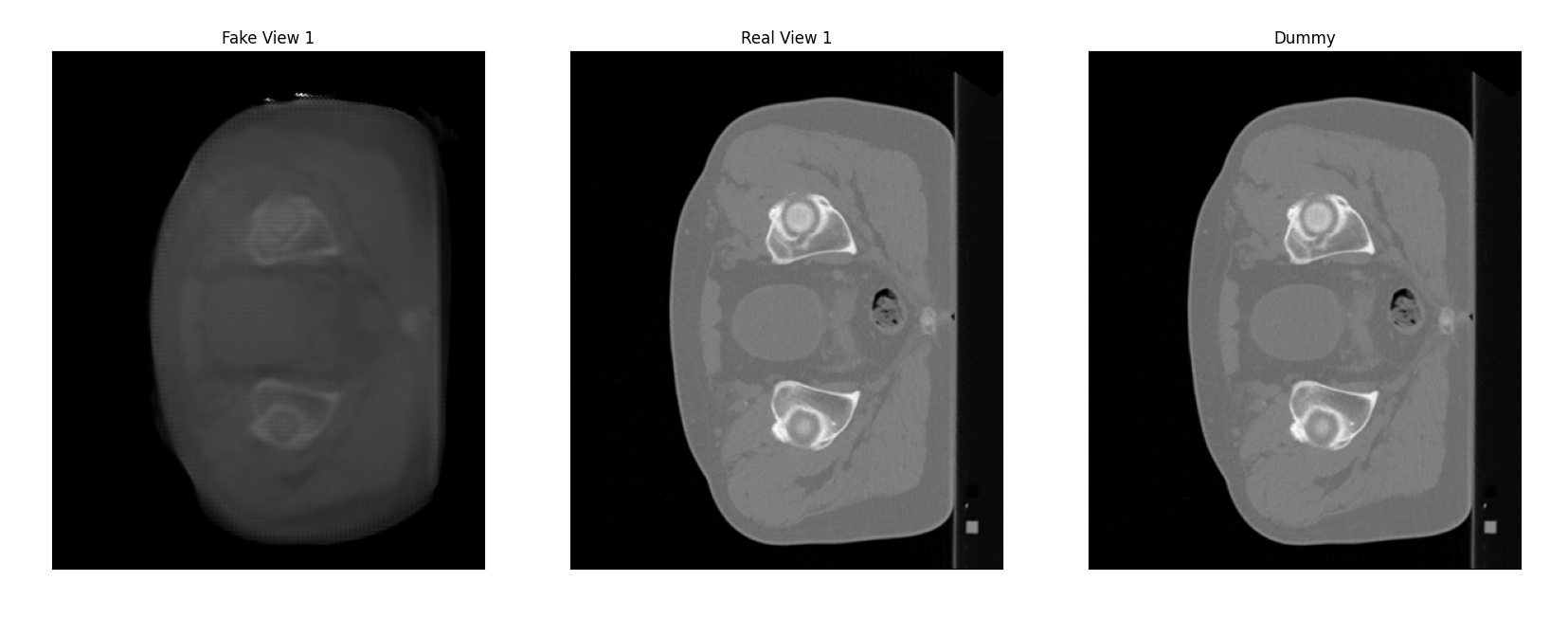}
    \caption{CycleGAN-1Res (Generated Image 3)}
    \label{fig:18}
\end{figure}

\begin{figure} [H]
    \centering
    \includegraphics[width=0.85\linewidth]{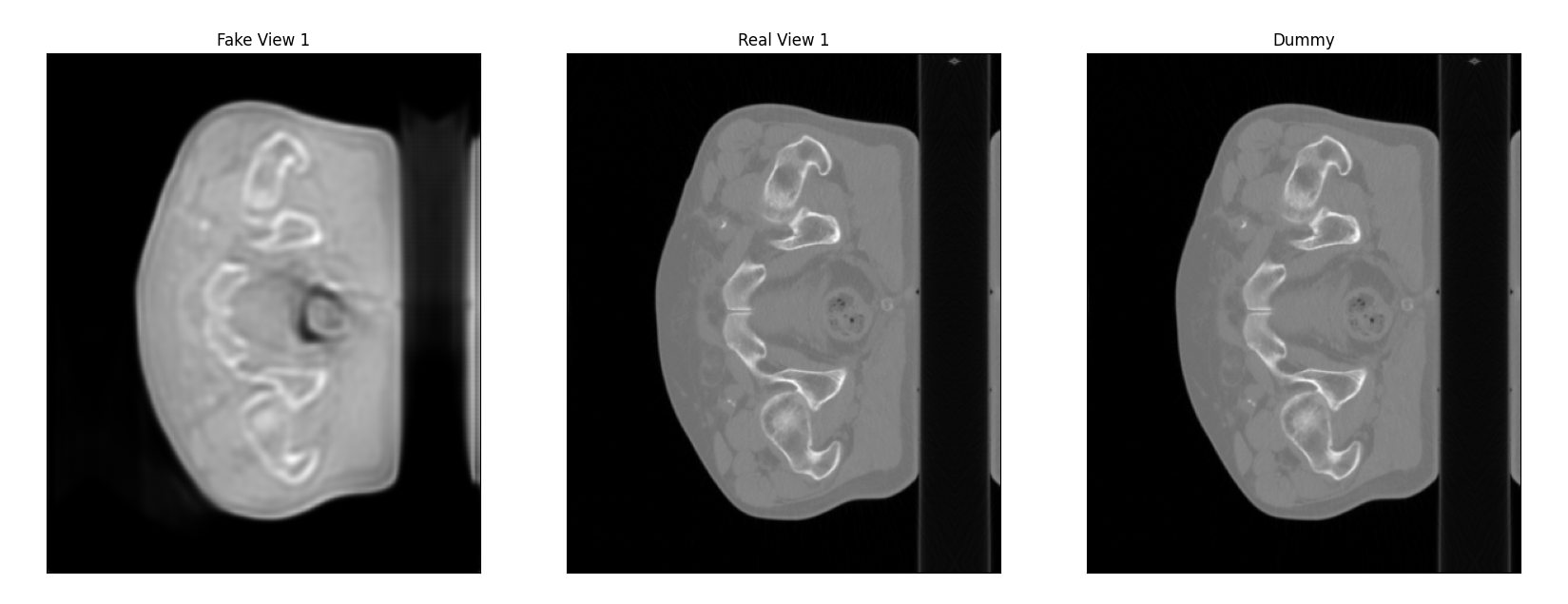}
    \caption{CycleGAN-Disc-20 (Generated Image 1)}
    \label{fig:19}
\end{figure}

\begin{figure} [H]
    \centering
    \includegraphics[width=0.85\linewidth]{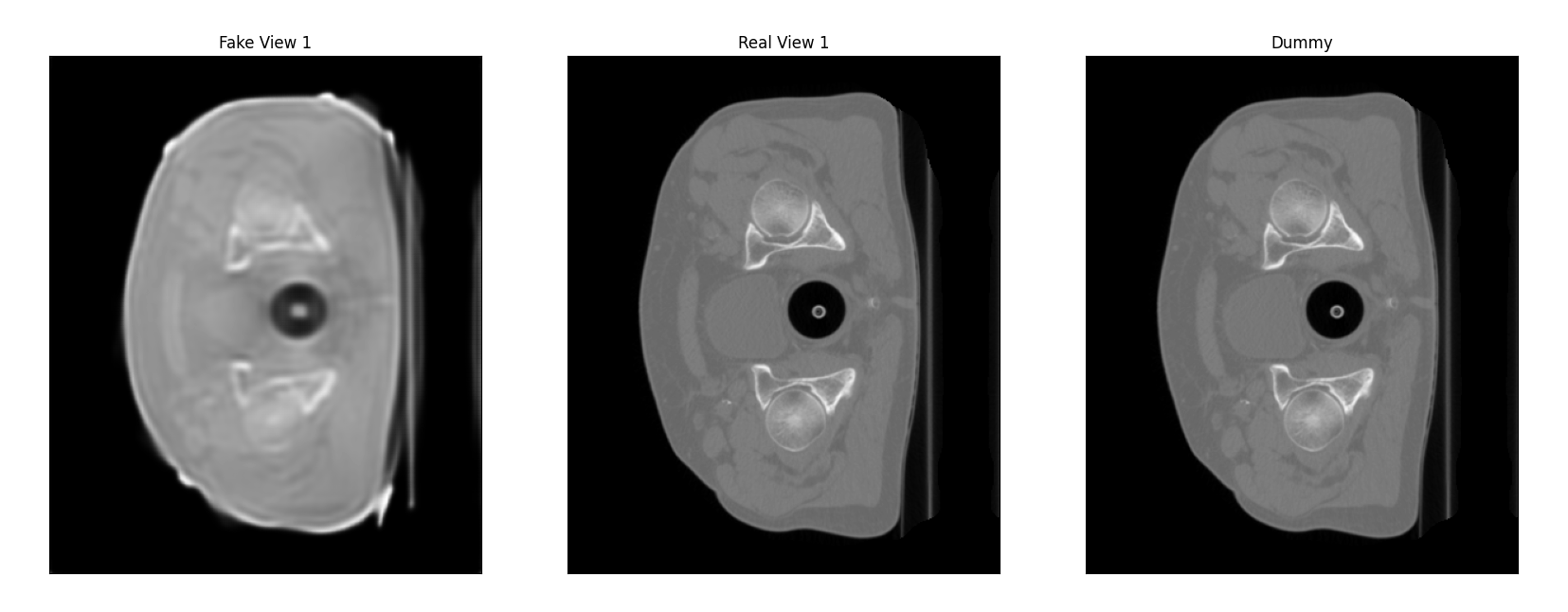}
    \caption{CycleGAN-Disc-20 (Generated Image 2)}
    \label{fig:20}
\end{figure}

\begin{figure} [H]
    \centering
    \includegraphics[width=0.85\linewidth]{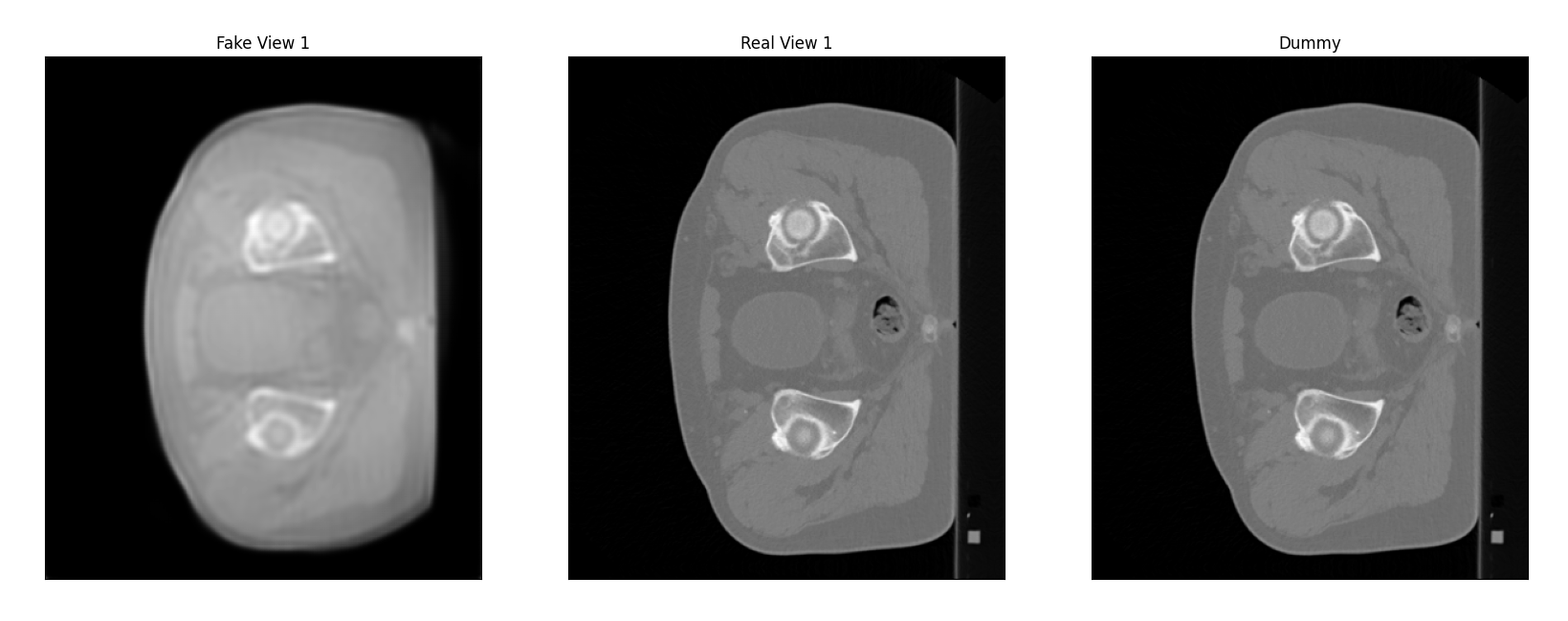}
    \caption{CycleGAN-Disc-20 (Generated Image 3)}
    \label{fig:21}
\end{figure}

\begin{figure} [H]
    \centering
    \includegraphics[width=0.85\linewidth]{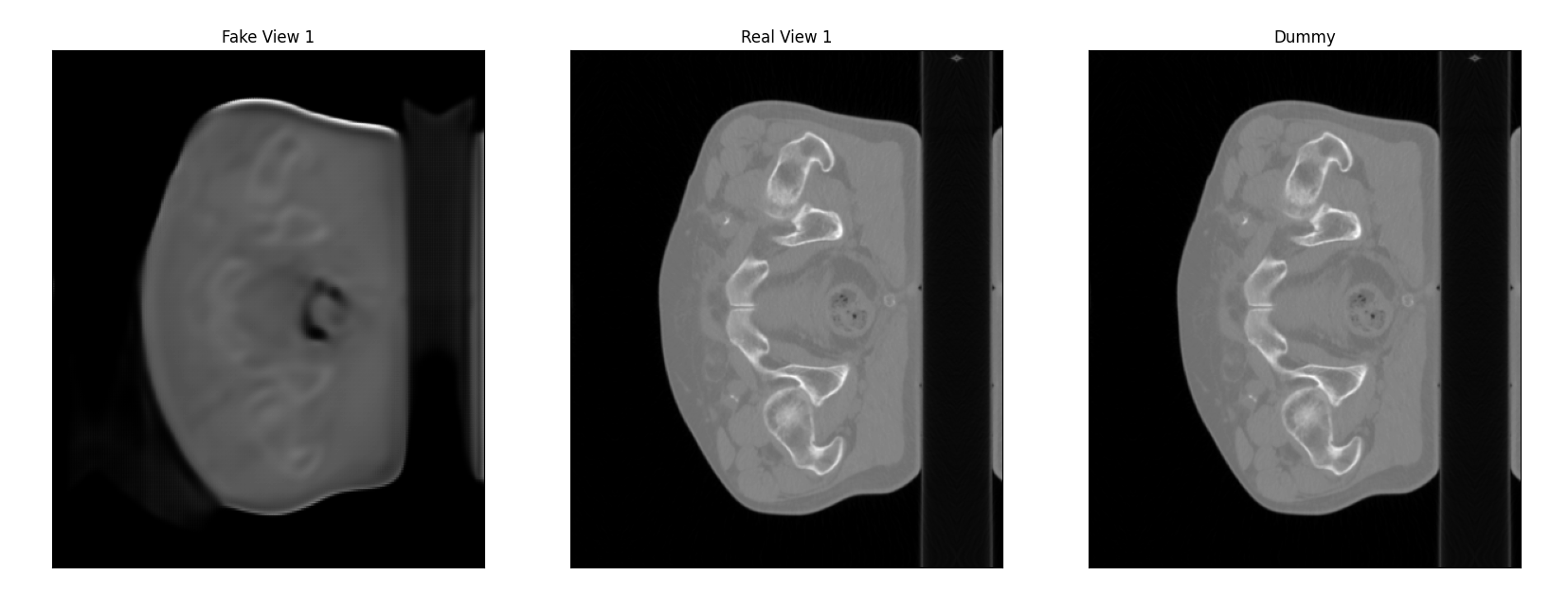}
    \caption{CycleGAN-Gen-20 (Generated Image 1)}
    \label{fig:22}
\end{figure}

\begin{figure} [H]
    \centering
    \includegraphics[width=0.85\linewidth]{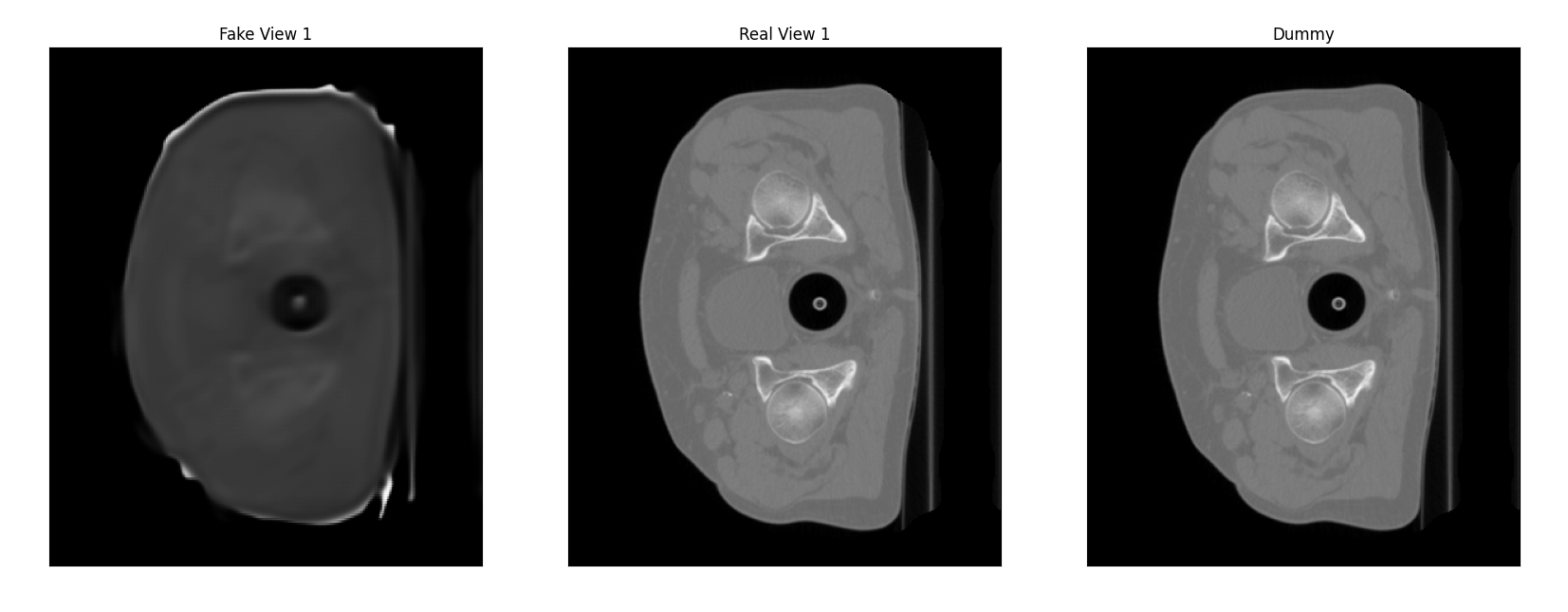}
    \caption{CycleGAN-Gen-20 (Generated Image 2)}
    \label{fig:23}
\end{figure}

\begin{figure} [H]
    \centering
    \includegraphics[width=0.85\linewidth]{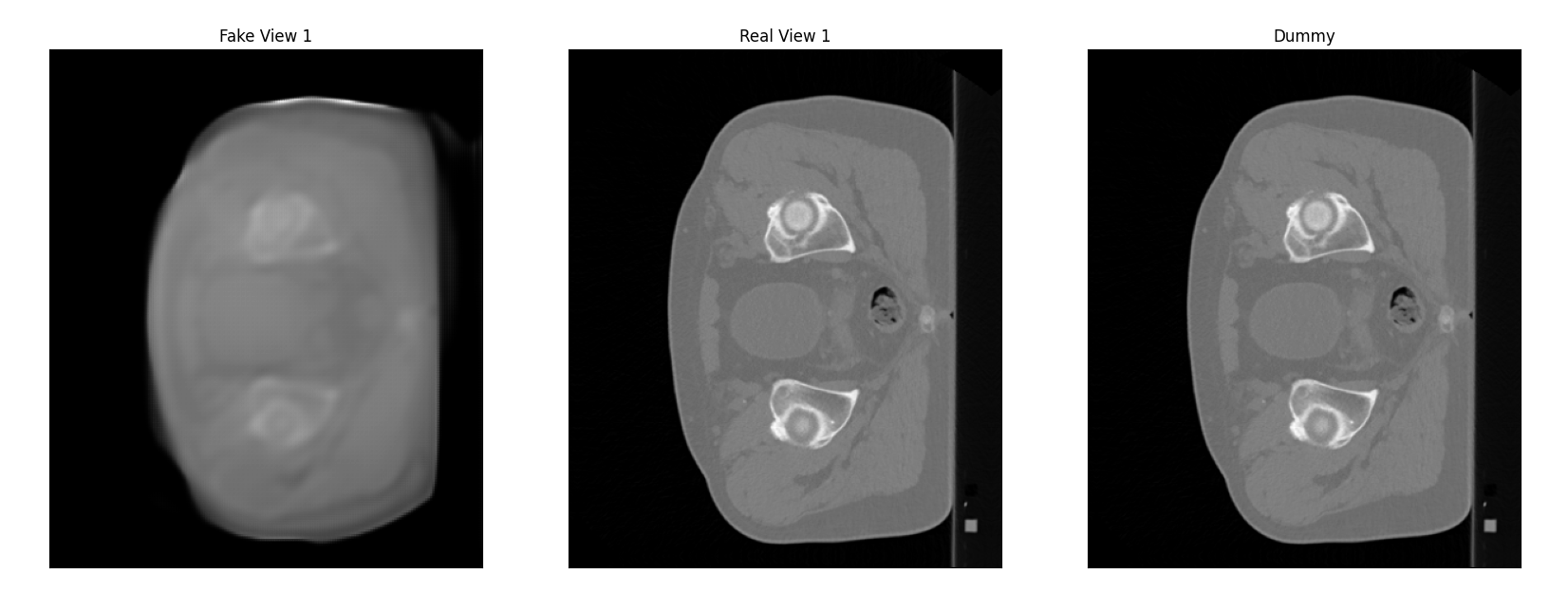}
    \caption{CycleGAN-Gen-20 (Generated Image 3)}
    \label{fig:24}
\end{figure}

\clearpage
\section{FQGA Performance with SEM Method}

\subsection{Quantitative Performance}

\begin{figure} [H]
    \centering
    \includegraphics[width=0.7\linewidth]{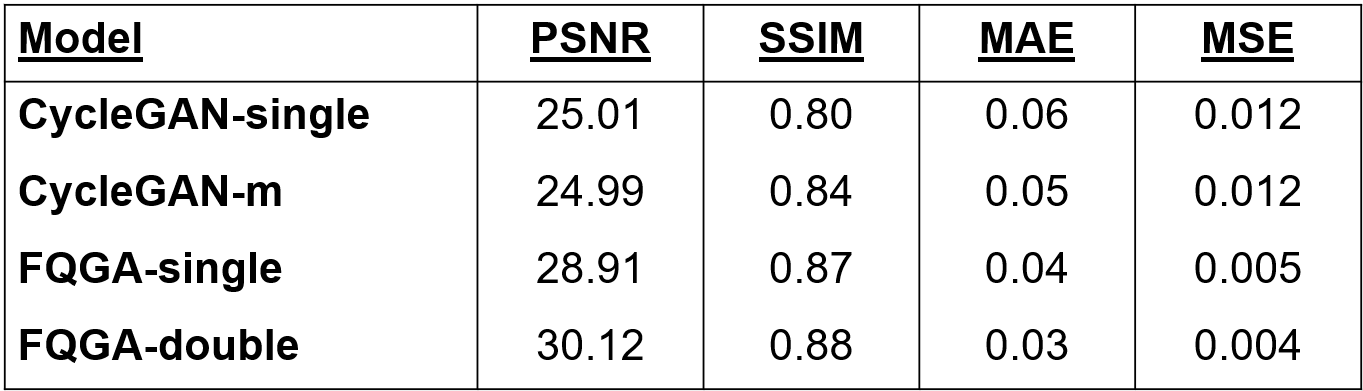}
    \caption{FQGA Performance with SEM Method}
    \label{fig:25}
\end{figure}

The results above are the models trained using the
Single-Epoch Modification (SEM). FQGA-single is FQGA model with 1 FQGA layer trained using SEM Method while FQGA-double is FQGA model with 2 FQGA-layers trained using SEM Method. CycleGAN-m
follows the implementation of Vanilla CycleGAN but
with filter sizes and the number of filters like those of
FQGA. With the additional FQGA layer, we can see
that the model performance of FGQA (double) - i.e. FQGA model with 2 FQGA-layers has
improved from FQGA (single) - i.e. FQGA model with 1 FQGA-layer which shows how
FQGA model was able to both cut down on the number
of parameters required while achieving superior
performance by extracting important compressed low level features while retaining an understanding of high level features in an image through concatenations and
skip connections.

\subsection{Quantitative Performance on LARGER TEST SET (100 SAMPLES – EXPERIMENT 2)}
Instead of 10 samples as tested in Section 9.1 of this
paper, We now present model performance on 100
paired volume data from SynthRAD dataset. A total of
100 paired volume data can be used for testing as there
are 150 paired volume data from SynthRAD dataset
minus 35 paired volume data for training data and
minus 15 paired volume data for validation data. In
Figure 28 below, we can also see that FQGA (single)
and FQGA (double) still outperforms or has comparable
performance with CycleGAN but with ¼ the number of
parameters as CycleGAN.

\begin{figure} [H]
    \centering
    \includegraphics[width=0.7\linewidth]{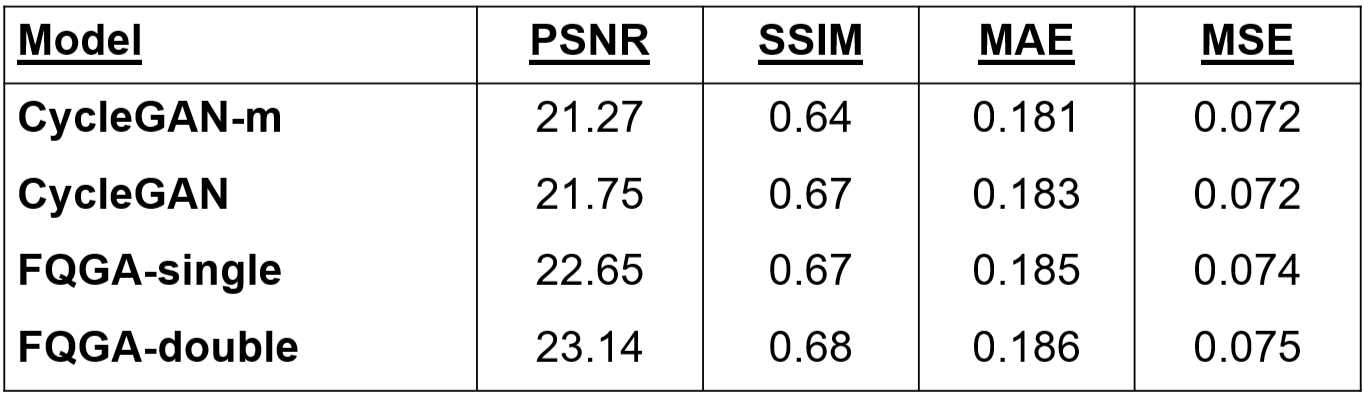}
    \caption{ Quantitative Test Performance (100 samples)}
    \label{fig:49}
\end{figure}

This shows FQGA-single and FQGA-double reliable
superior performance in being able to outperform
CycleGAN-m and CycleGAN. Therefore, this very
much reduces the possibility of FQGA model, and its
variants (FQGA-single and FQGA-double) better
performance being attributed to chance.

\clearpage
\subsection{Qualitative Performance}

\textbf{*Images From Top to Bottom – FQGA (single)
generated sCT, CycleGAN generated sCT, Original real
CT}

\begin{figure} [H]
    \centering
    \includegraphics[width=0.7\linewidth]{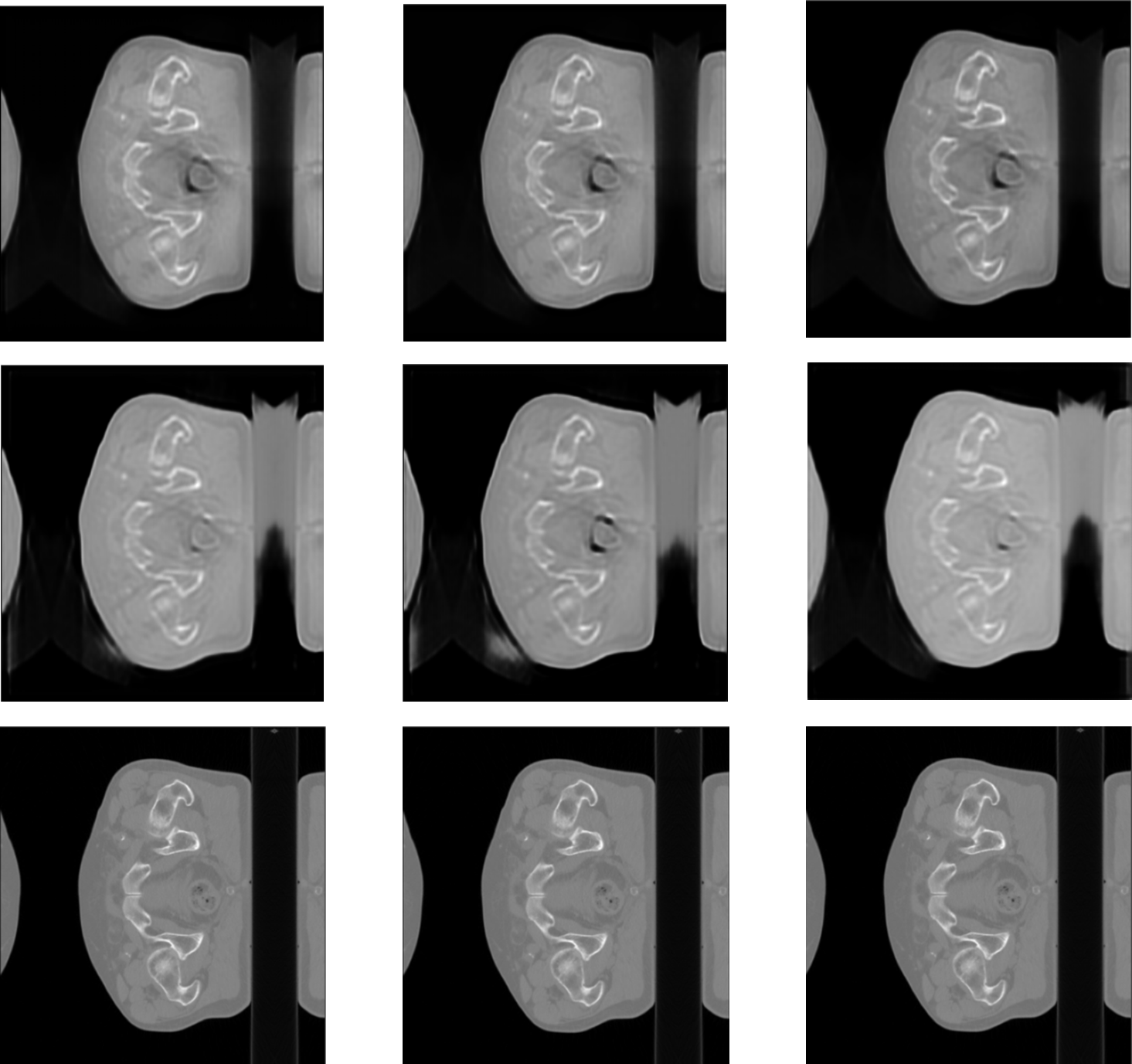}
    \caption{FQGA Qualitative Performance}
    \label{fig:26}
\end{figure}

\subsection{Comparison of Tissue HU Value Distribution (FQGA vs CycleGAN)}

\begin{figure} [H]
    \centering
    \includegraphics[width=\linewidth]{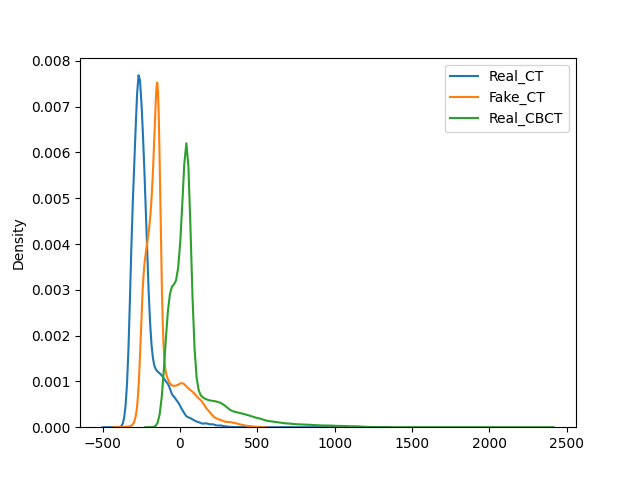}
    \caption{FQGA-single KDE Plots}
    \label{fig:27}
\end{figure}

\begin{figure} [H]
    \centering
    \includegraphics[width=\linewidth]{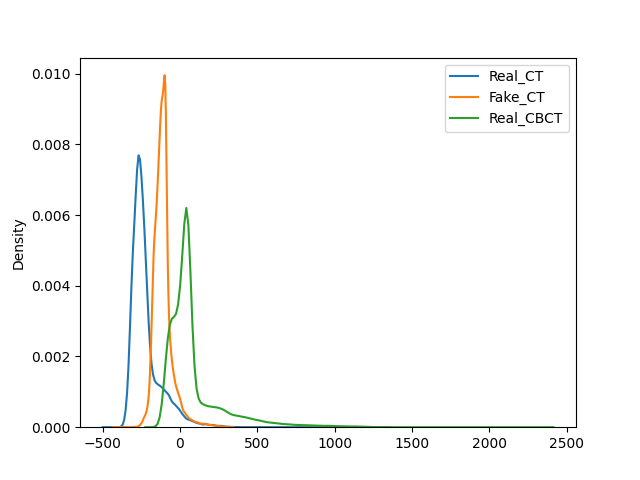}
    \caption{CycleGAN KDE Plots}
    \label{fig:28}
\end{figure}

From Section 9.1 and 9.2 (Quantitative Performance), we can
infer from its quantitative performance to draw
conclusions about its HU value distribution, that is, on
aggregate, the HU value distribution of sCT images
generated by FQGA models are closer to the true
underlying HU value distribution of CT images
compared to those produced by CycleGAN model. From Section 9.3, we can see that the sCT images generated by FQGA produced less image noise artifacts and smudges compared to CycleGAN.
In Section 9.4, we now look at 2 specific HU value distribution
examples of specific tissues within the sCT image as
seen in Figure 30 and Figure 31. From Figure 30,
we can see that the generated HU value distribution of
sCT generated by FQGA (orange-line) seems to resemble
more closely to the true HU distribution of the CT
image (blue-line) compared to the generated HU value
distribution of sCT generated by CycleGAN (orange-line)
in Figure 31.

\clearpage
\section{Conclusion}
This paper has introduced a novel Image-to-Image Translation model called FQGA model. Through the Ablation Studies of FQGA in Section 8, we noticed that FQGA’s Discriminator Model was a huge factor which led to the significant improvement in model’s performance both
quantitatively and qualitatively. However, with the
introduction of FQGA’s Generator Model, this also
offered a slight improvement in the PSNR score.

An exploratory tour into models trained after a single epoch via SEM (Single-Epoch Modification) Method showed astonishing results. SEM method is used for the training of 3D volume data, in this case, medical patient data which gives rise to the models being trained on a single-epoch vs multi-epoch variations. In single-epoch, each patient data in the dataset is only passed through the model once. CycleGAN-single which was implemented using SEM, although has its quantitative results improved, qualitative visual results were still poor. However, when compared to FQGA model performance, even after training on a single epoch, FQGA showed impressive results.

In the future, more research can be done to see how FQGA performs in other Image-to-Image translation tasks in other non-medical domains. Also, more research can be done to understand whether the FQGA’s Generator Model increased the robustness of the overall model. These performance gains even with fewer model parameters and fewer epochs (which will result in time
and computational savings) may be applicable to other image-to-image translation tasks in Machine Learning apart from the ones discussed in this paper, opening many exciting avenues for real-world applications.

Lastly, the Appendix Section of this paper hopes to
supplement certain points in this paper although it is
possible that certain points may still have room for
improvement. In the Appendix, a more comprehensive
argument of the SEM method, Single-Epoch Models
and FQGA model were highlighted in this paper for the
curious reader.

\clearpage
\section*{Acknowledgments}

I would like to express my greatest gratitude to my
supervisor, Professor Cai Yiyu. I am grateful to be able
to work on a project of my interest under the expertise and supervision of Professor Cai.

Next, I would also like to extend my appreciation and
gratitude to Dr. Nei Wen Long from SingHealth –
Singapore General Hospital (SGH) and Dr. Tan Hong Qi
from SingHealth – National Cancer Centre Singapore
(NCCS) for sharing their valuable feedback and
thoughts throughout the research journey. In addition,
their medical domain knowledge, expertise, experience,
patience and technical guidance are extremely valuable
in helping to understand the problem statement and
aided in re-evaluating the research direction for it to be
more meaningful and impactful.

Lastly, I would also like to express my thanks to Research
Associate, Dr. Azam Abu Bakr for helping with
administrative access to the GPU servers and his
debugging efforts to resolved process bugs. All his
effort facilitated the ease of use of remote GPUs and
allowed for parallel processing across multiple GPUs to
be performed successfully.

\textbf{All papers arising as a result of URECA programme must have the following (important) statement under the acknowledgement section.}

\textit{I would like to acknowledge the funding support from Nanyang Technological University – URECA Undergraduate Research Programme for this research project.}

\textbf{Acknowledgement also applies to Journal papers, Conference papers and Proceedings of URECA Undergraduate Research papers arising as a result of URECA.}




\clearpage
\section*{Appendix}

\section{Dataset Exploration}
For data cleaning in Section 3.1, an iterative data
exploration process of the SynthRAD dataset is
performed. For each training batch, we used a reference
3D CBCT-CT paired data which is a held-out volume to
verify the 3-way KDE plots are produced and generated
by the model after every successful training of all the
2D paired CBCT-CT slices of a single patient right 
before moving on to the next patient. This algorithm
showed when KDE plot predictions deviate from that
training batch. Anomalous pairs are then repaired until
consistent with reference KDE plot. This was a
necessary step as some of the CBCT-CT paired data
were paired inconsistently and in reverse etc. Doing so
allowed the correct extraction of CBCT and CT slice
data from the CBCT-CT paired data. Also, this
algorithm was crucial in identifying some anomalous
slices which seemed to be introducing noise to the
dataset which affected model performance. Therefore,
when plot fails to match reference even after re-pairing,
they are deemed “faulty” and discarded.

\section{CYCLEGAN 200 EPOCHS LOSS}

\begin{figure} [H]
    \centering
    \includegraphics[width=\linewidth]{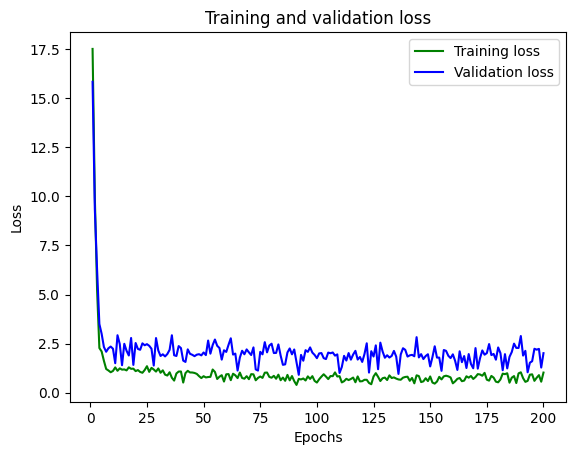}
    \caption{Generator Loss from CT to CBCT}
    \label{fig:39}
\end{figure}

\begin{figure} [H]
    \centering
    \includegraphics[width=\linewidth]{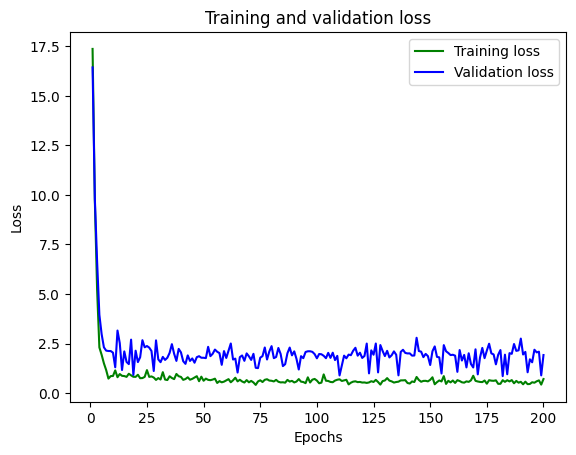}
    \caption{Generator Loss from CBCT to CT}
    \label{fig:40}
\end{figure}

\begin{figure} [H]
    \centering
    \includegraphics[width=\linewidth]{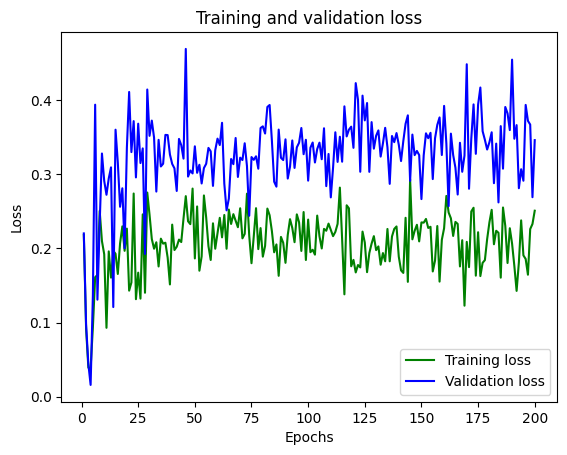}
    \caption{Discriminator Loss for CT}
    \label{fig:41}
\end{figure}

\begin{figure} [H]
    \centering
    \includegraphics[width=\linewidth]{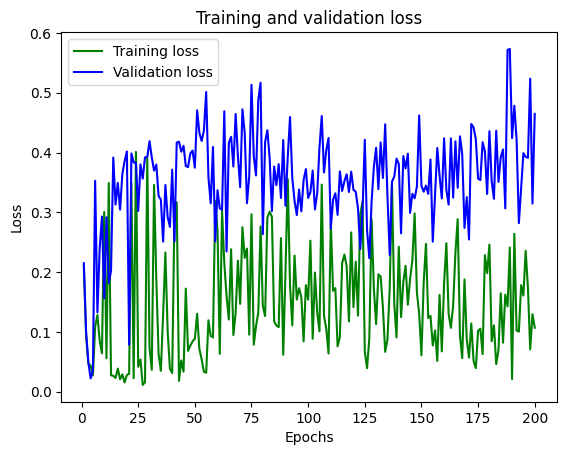}
    \caption{Discriminator Loss for CBCT}
    \label{fig:42}
\end{figure}

\begin{figure} [H]
    \centering
    \includegraphics[width=\linewidth]{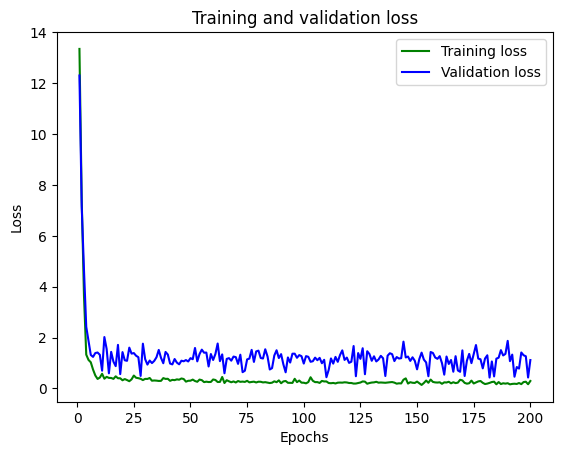}
    \caption{Cycle Loss}
    \label{fig:43}
\end{figure}

\section{CYCLEGAN-1RES 200 EPOCHS LOSS}

\begin{figure} [H]
    \centering
    \includegraphics[width=\linewidth]{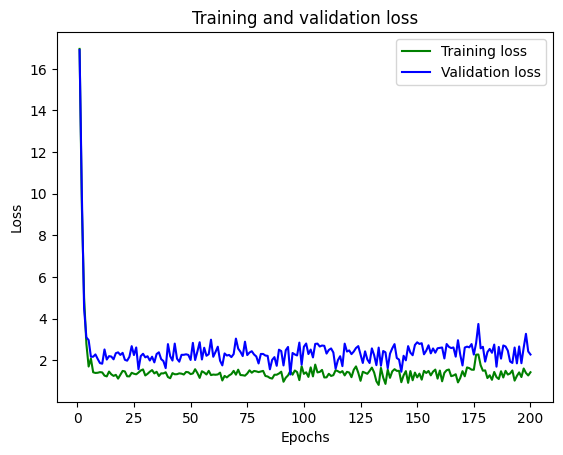}
    \caption{Generator Loss from CT to CBCT}
    \label{fig:44}
\end{figure}

\begin{figure} [H]
    \centering
    \includegraphics[width=\linewidth]{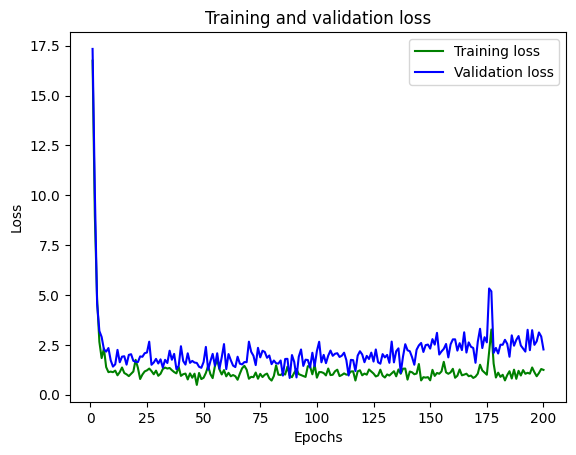}
    \caption{Generator Loss from CBCT to CT}
    \label{fig:45}
\end{figure}

\begin{figure} [H]
    \centering
    \includegraphics[width=\linewidth]{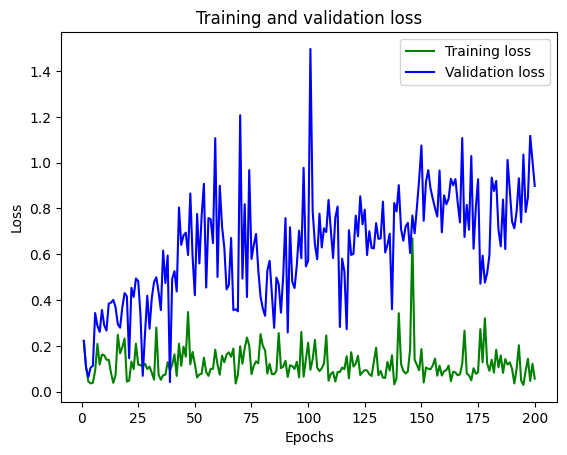}
    \caption{Discriminator Loss for CT}
    \label{fig:46}
\end{figure}

\begin{figure} [H]
    \centering
    \includegraphics[width=\linewidth]{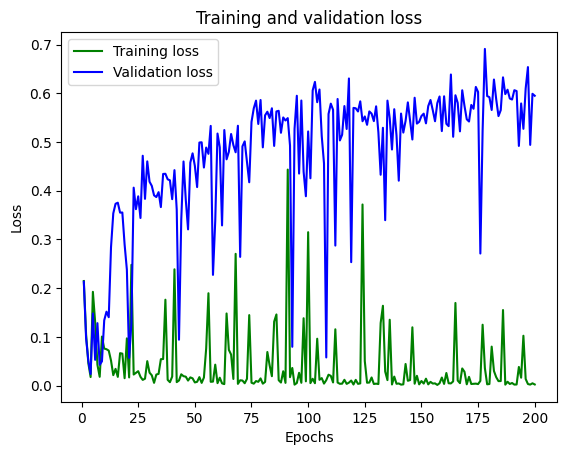}
    \caption{Discriminator Loss for CBCT}
    \label{fig:47}
\end{figure}

\begin{figure} [H]
    \centering
    \includegraphics[width=\linewidth]{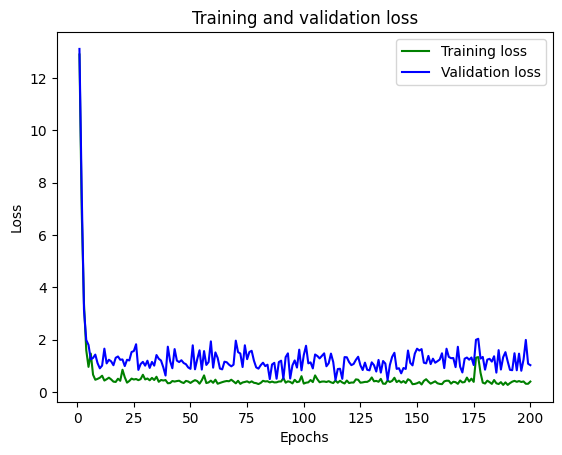}
    \caption{Cycle Loss}
    \label{fig:48}
\end{figure}

\section{Single-Epoch Modification (SEM) Method}
\subsection{SEM Overview}
In usual training of paired data for image-to-image
translation problems, all 3D CBCT-CT volume pair
data for all patients are converted into 2D CBCT-CT
slice pair data and shuffled across different patients
altogether. I hypothesize that doing so may cause the
model to miss out on learning crucial relationships
between slices of the same patient as each patient may
have unique characteristics like HU distribution. Instead
of training on all the patient data at once and shuffling
data across all patients, single-epoch modification
(SEM) here refers to training on one patient data at a
time before moving to the next patient. In more detail,
this means that the CBCT-CT paired 2D data are
shuffled only within that patient and after the model has
finished training on one patient, its weights and
optimizers are saved and loaded into a model before the
training of the next patient. For example, given we have
35 patients or 35 CBCT-CT paired volume data for
training, after one epoch, we will have 35 model
weights and model optimizers being saved where the 1st
model weights and optimizers represent the model’s
weights and optimizers after training on the 1st patient.
When training continues, the 1st patient model’s weights
and optimizers will be first loaded into the model before
training the next patient, that is the 2nd patient. When
going on to the next epoch, the model will retrieve the
model’s weights and optimizers saved from the last
patient (e.g. 35th patient) from the previous epoch.
Results from training the model in this way produced
better quantitative and qualitative performance in faster
time which may be due to loss landscape being
smoother and easier to optimized by Adam optimizer as
when training on one patient at a time, the amount of
noise in the data is contributed only by 1 patient instead
of multiple patients which also is an easier challenge for
the algorithm to be invariant to this noise instead of the combination of multiple noise artifacts from different
patients. Lastly, training the model in this way lets the
model be more robust as in a way, it averages the model
weights across the different patients while at the same
time, learning richer mappings and relationships not just
between CBCT and CT images but also between 2D CT
or CBCT images which are on top and below each other
in the depth-dimension of the 3D volume data.

\subsection{Identifying Best Single Epoch Model}
Training a model on one epoch substantially improves
the diversity of samples processed by the model over
the course of training. Training for E epochs is roughly
equivalent to training on a shuffled dataset consisting of
E copies of the original dataset for one epoch. This
means that the diversity of the original dataset is E
times less than that of the one epoch training. \cite{Komatsuzaki}
Diversity in dataset is important to improve the
performance of models \cite{Hestness}, \cite{Radford}.
Although in this paper this problem statement of
converting between CBCT and CT images is trained
using supervised image-to-image translation paired
training data, a similar paper on unsupervised learning
regarding single epoch training is discussed in 2019 \cite{Komatsuzaki}.
“One Epoch Is All You Need” suggests training on a
larger dataset for only one epoch unlike the current
practice, where models are trained from tens to
hundreds of epochs. The paper result suggests that we
try to make the ratio of the number of “processed
tokens” over the number of parameters, or T/P, as close
to 5 as possible. This result suggests that five words per
parameter can be the most efficiently compressed, at
least in their setting \cite{Komatsuzaki}. T=cI, where c is the number of
“tokens” per minibatch, and I is the number of
iterations. The model with size P is trained for I
iterations for one epoch without any regularization
method \cite{Komatsuzaki}. In the paper, “One Epoch Is All You Need”
relates to the training of Language Models with tokens
used in its data pipeline. In contrast, for our problem of
image-to-image translation between CBCT and CT
images, this is a Computer Vision Generative task, and
we used regularization in the form of Dropout.

\begin{figure} [H]
    \centering
    \includegraphics[width=0.5\linewidth]{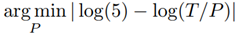}
    \caption{Single Epoch Model Heuristic \cite{Komatsuzaki}}
    \label{fig:29}
\end{figure}

According to Figure 42, we set P according to this
heuristic and set the ratio T/P as close to 5 as possible
or equivalently solving the equation in Figure 42.
Instead of tokens, we let the tokens in the training of
Language Models be replaced by pixels of image in our
problem case. In our case, the resolution of images is
436 pixels by 416 pixels. The Generator Model in
FQGA has about P equals to 4 million parameters. As
we set I, the number of iterations to be 1, and T=cI, and
given the number of slices ranges from 50 to 105
(Section 1.1) per patient, T ranges from 9 million to 19
million. Therefore, T/P = [2.25, 4.75] which means that
in this context, 2.25 pixels to 4.75 pixels per parameter
can be the most efficiently compressed.

The adoption of the ideas from the paper “One Epoch Is
All You Need” for a computer vision, Image-To-Image Translation
task is not surprising as even in their paper,
it mentions how Regularization methods are crucial for
computer vision tasks and may benefit from one epoch
training even more. According to their paper, a
test/validation dataset is not crucial as averaging the
train loss per minibatch measured on the past n
minibatches is approximately equals to the test loss of n
is small enough \cite{Komatsuzaki}. In our paper, we still will have a
validation and test dataset to assess the performance of
the model.

\subsection{STABILIZATION OF TRAINING LOSS}
\textbf{*(DOMAIN A – CT, DOMAIN B – CBCT)}

In this experiment, 1 epoch consists of training through
35 CBCT-CT volume data pairs. Therefore, the
following plots below will show the generator losses for
the respective models. As seen, $gen_AtoB_loss$ refers to
generator loss for translating images from CT (A)
domain to CBCT (B) domain while $gen_BtoA_loss$
refers to the generator loss for translating images from
CBCT (B) domain to CT (A) domain.

\begin{figure} [H]
    \centering
    \includegraphics[width=\linewidth]{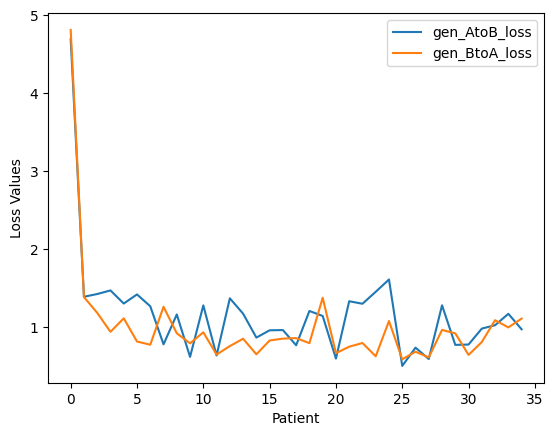}
    \caption{CycleGAN Training Loss after 35 patients}
    \label{fig:30}
\end{figure}

\begin{figure} [H]
    \centering
    \includegraphics[width=\linewidth]{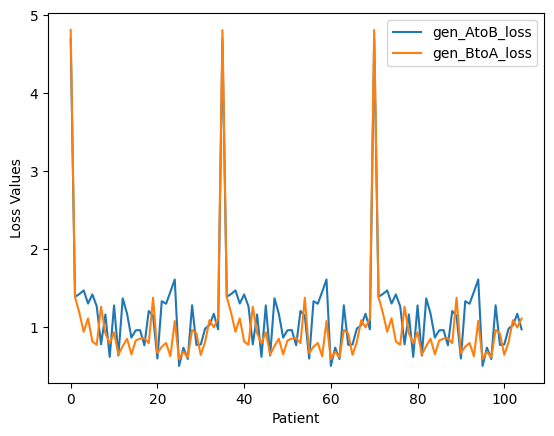}
    \caption{CycleGAN Training Loss after training on
the same 35 patients 3 times}
    \label{fig:31}
\end{figure}

However, for CycleGAN it can be seen that saturation
is not as smooth. However, if we try to train for more
than 3 epochs (Figure 44), this similar pattern repeats
and no smoothening of loss function occurs. Moreover,
Quantitative validation performance on metrics like
PSNR and SSIM falls with more training epochs and
qualititative visual results deteoriate as seen in Section 14.5, Figure 51. Therefore, we consider Figure 43 as the
saturation of CycleGAN loss function.

Now that we have determined that CycleGAN
Generator models have attained plateau in their training
loss functions during the first epoch, we now have to
determine after which patient is model performance
most optimal which we will determine in Section 14.4.

\subsection{VALIDATION PERFORMANCE (QUANTITATIVE)}
After Section 14.3 where we verify that the training
losses of models have in fact saturated, we are now
going to do evaluate models on validation data to
identify the best model weights for CycleGAN.
Validation is performed on 15 CBCT-CT volume data
pairs validation data which was not used during the
training process of models. In Section 14.3, Figure 43, we see that
plateau of loss function occurs from patient 15 onwards.
Therefore, we took the models saved after training on
the 15th patient until the model saved after training on
the 35th patient to evaluate on the validation data of 15
volume pairs.

\begin{figure} [H]
    \centering
    \includegraphics[width=\linewidth]{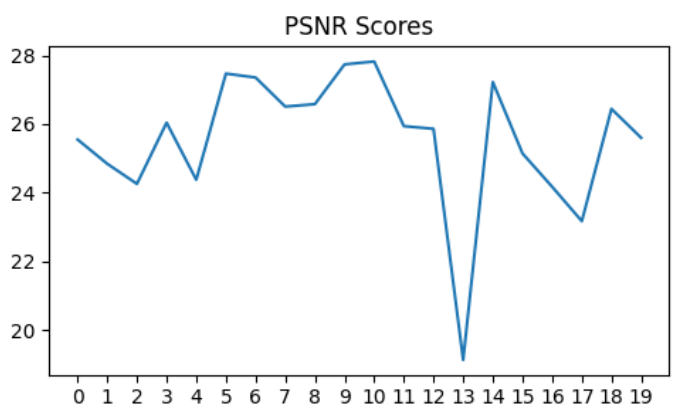}
    \caption{CycleGAN – PSNR Validation}
    \label{fig:32}
\end{figure}

\begin{figure} [H]
    \centering
    \includegraphics[width=\linewidth]{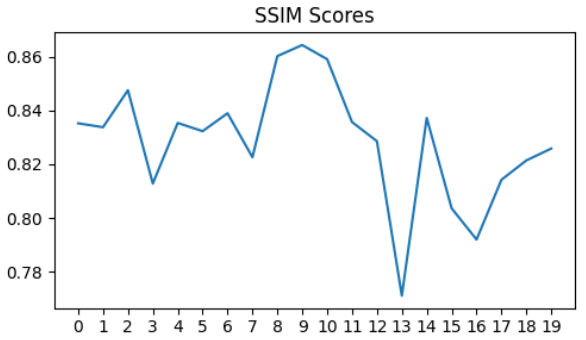}
    \caption{CycleGAN – SSIM Validation}
    \label{fig:33}
\end{figure}

\begin{figure} [H]
    \centering
    \includegraphics[width=\linewidth]{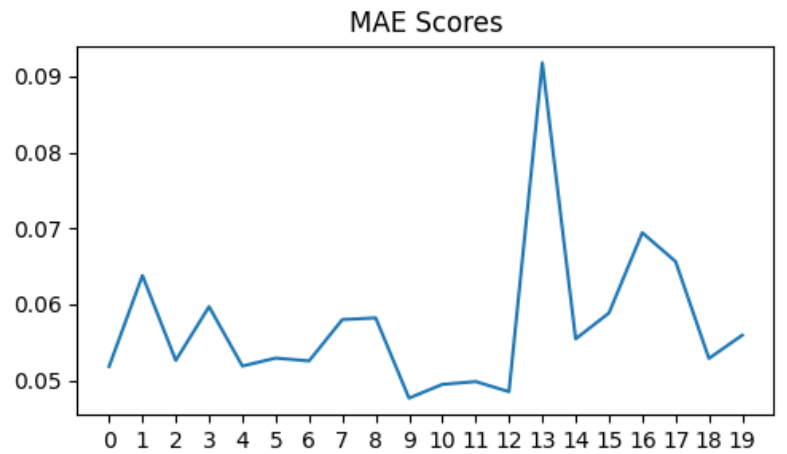}
    \caption{CycleGAN – MAE Validation}
    \label{fig:34}
\end{figure}

\begin{figure} [H]
    \centering
    \includegraphics[width=\linewidth]{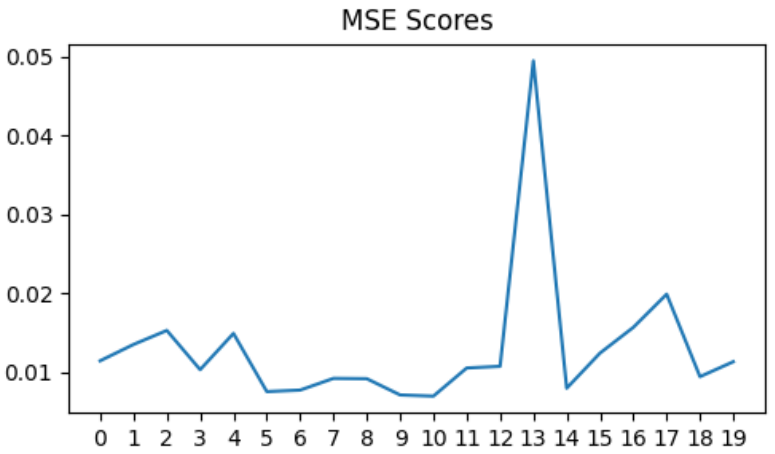}
    \caption{CycleGAN – MSE Validation}
    \label{fig:35}
\end{figure}

In the above Figure 45, 46, 47, 48, the 0 and
19 on the x-axis indicates the model performance after
training on the 15th patient and 35th patient respectively.
Based on the above quantitative validation performance
plots, For FQGA (single) we can see that optimal
quantitative model is after the training of the 31st patient
while optimal CycleGAN model is after the training of
the 30th patient. We further justify this in Section 14.5 via
qualitative validation performance.

\subsection{VALIDATION PERFORMANCE (QUALITATIVE)}
In the previous Section, Section 14.4, we have determined from a Quantitative perspective that
after the training of the 31st patient, optimal performance is attained. In Section 14.5, we verify this further.
In Section 14.3, Figure 43, the training loss function of our models
drops drastically only after training of a few patients.
This fast learning of the model to be able to find a
mapping function between CBCT and CT images is
reinforced by the high SSIM and PSNR test scores
when model was evaluated at those points. For
example, FQGA (single) model after the training of 9th
patient had validation scores of 27.88, 0.84, 0.05, 0.006
for PSNR, SSIM, MAE, MSE respectively. However,
visual qualitative results were compromised during
these stages as seen in Figure 49 compared to Figure 50 which shows the results after training of the 31st
patient. This may mean a local minimum may have been
obtained in Figure 49 instead of a global one.

\begin{figure} [H]
    \centering
    \includegraphics[width=0.85\linewidth]{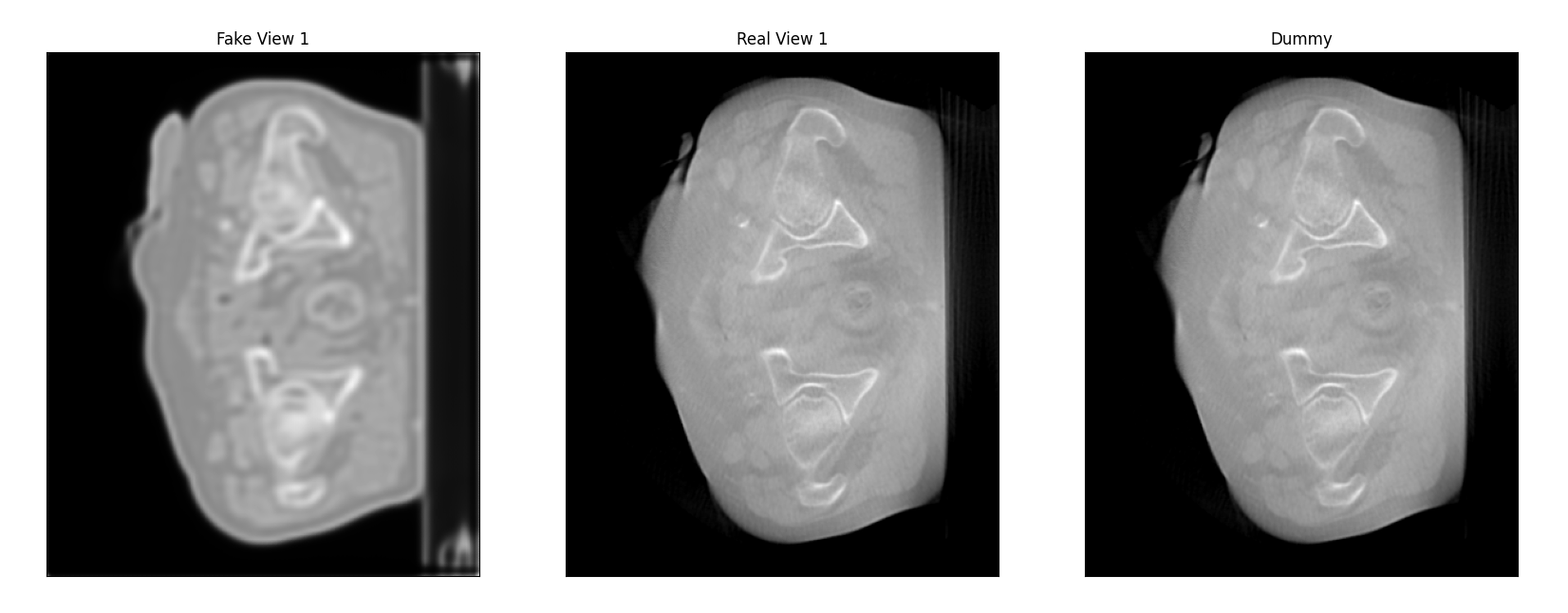}
    \caption{Generated sCT after 9th patient}
    \label{fig:36}
\end{figure}

\begin{figure} [H]
    \centering
    \includegraphics[width=0.85\linewidth]{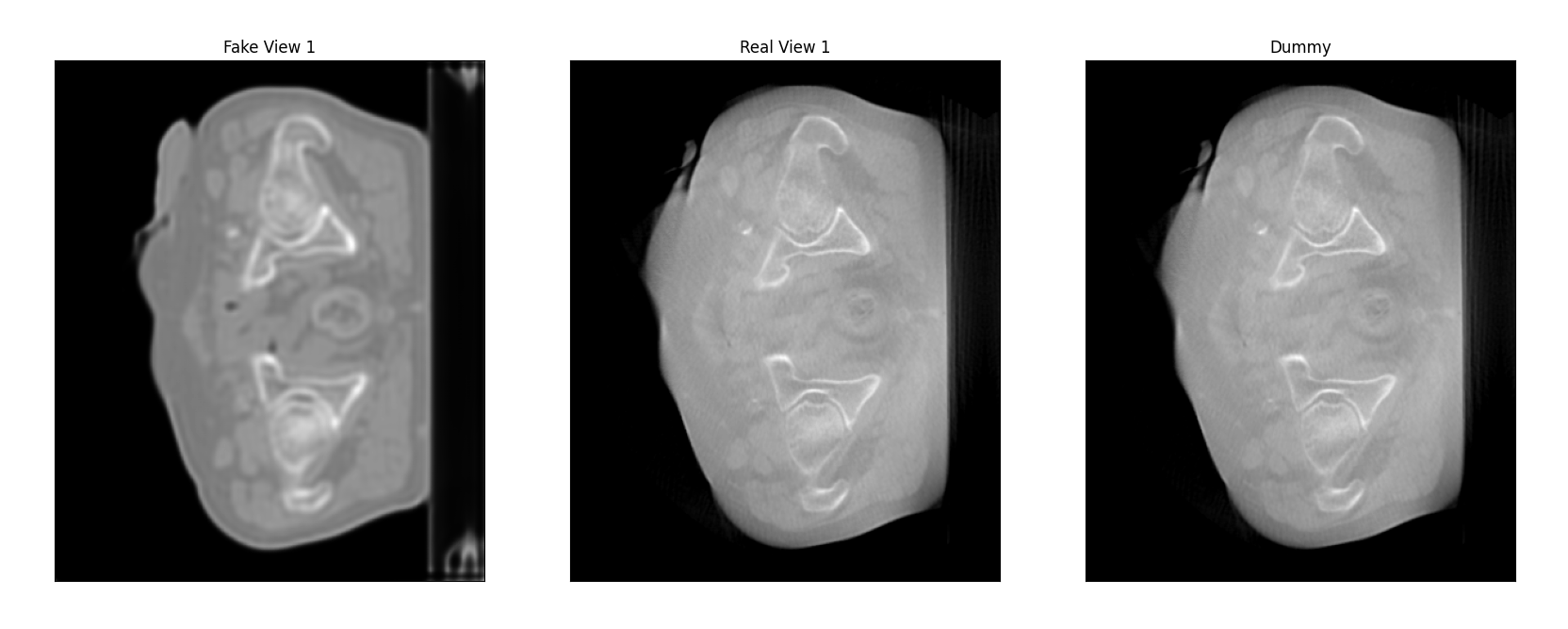}
    \caption{Generated sCT after 31st patient}
    \label{fig:37}
\end{figure}

So far, for the SEM Method, we retrieved the model that gave optimal performance from the procedure described in Section 14, after training on a certain number of patients to generate visual images for inspection.
More training is not beneficial, and in fact, detrimental as from
a qualitative visual perspective, with further training,
the images became less clear as seen in Figure 51.

\begin{figure} [H]
    \centering
    \includegraphics[width=0.85\linewidth]{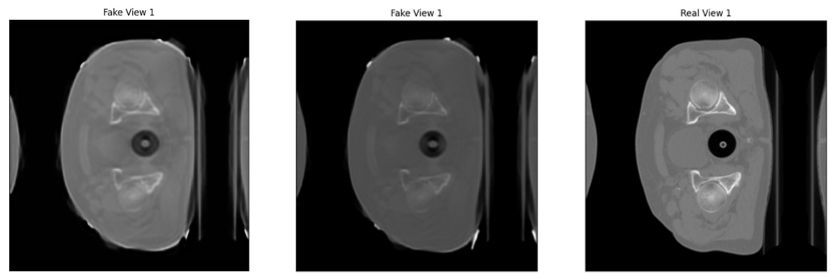}
    \caption{Left – Enough training,
Middle – Over training, Right – Real CT image}
    \label{fig:38}
\end{figure}

Therefore, in Section 14, so far, we explained how optimal single
epoch models for FQGA (single) and CycleGAN were
obtained, we now repeat the same steps for FQGA
(double) and CycleGAN-m (modified).

\subsection{Final Comments ON SINGLE-EPOCH MODIFICATION (SEM) Method}

\subsubsection{Variations in Training of Patients}
Instead of training each patient once, we tried training
on each patient twice and saving the model weights and
optimizers for both its first- and second-time training on
the patient. But when training on the next patient, we
still loaded the weights after the first training on the
patient and not the second. However, when we
evaluated the test performance of the model after being
trained two times on the same patient, there is an
improve in model performance. For example, for FQGA
(single) model test performance of PSNR increased
from 29.12 to 29.74, SSIM increased from 0.873 to
0.876, MAE falls from 0.043 to 0.038 and MSE falls
from 0.0050 to 0.0043. Although marginal performance
gains, this is still an interesting avenue for further
research.

\subsubsection{More FQGA Layers}
FQGA (triple) with 3 FQGA layers was also trained,
but its performance dropped from FQGA (double)
which may mean that 2 FQGA layers may be the
optimal number of FQGA layers.


\end{document}